\documentclass[a4paper,fleqn]{cas-sc}

\usepackage[numbers,sort&compress]{natbib}

\usepackage{algorithm,algorithmic}
\usepackage{ragged2e}
\usepackage{amsmath,amssymb,amsfonts}
\makeatletter
\@fleqntrue
\makeatother
\usepackage{graphicx}
\usepackage{hyperref}
\usepackage{textcomp}
\usepackage[none]{hyphenat}
\usepackage{graphicx}
\usepackage{multirow}

\usepackage{enumitem}
\usepackage{tikz}
\usepackage{lipsum}
\usepackage{multicol}
\usepackage{float}
\usepackage{pifont}
\usepackage{makecell}

\usepackage{pgfplots}
\pgfplotsset{compat=1.18}
\usepackage{graphicx}
\usepackage{orcidlink}
\usepackage{pgfplotstable}
\usetikzlibrary{arrows.meta}
\usepackage{graphicx}
\usepackage{epstopdf}
\usepackage{silence}
\usepackage{microtype}

\usepackage{caption}
\let\oldalgorithmic\algorithmic
\renewcommand{\algorithmic}{\normalsize\oldalgorithmic}
\usepackage{float}
\usepackage[capitalize,nameinlink]{cleveref}

\def\tsc#1{\csdef{#1}{\textsc{\lowercase{#1}}\xspace}}
\tsc{WGM}
\tsc{QE}
\tsc{EP}
\tsc{PMS}
\tsc{BEC}
\tsc{DE}

\begin{document}
\sloppy
\let\WriteBookmarks\relax
\def\floatpagepagefraction{1}
\def\textpagefraction{.001}

\shorttitle{VesselSAM}

\shortauthors{Adnan Iltaf et~al.}

\title [mode = title]{VesselSAM: Leveraging SAM for Aortic Vessel Segmentation with AtrousLoRA}

%
\author[1,2]{Adnan Iltaf}[style=chinese]
\ead{adnan@siat.ac.cn}

\affiliation[1]{organization={Shenzhen Institute of Advanced Technology,
Chinese Academy of Sciences},
    city={Shenzhen},
    postcode={518055}, 
    country={China}}

\affiliation[2]{organization={University Chinese Academy of Sciences},
    city={Beijing},
    postcode={101408}, 
    country={China}}

\author[1]{Rayan Merghani Ahmed}[style=chinese]
\ead{rayan@siat.ac.cn}

\author[1]{Zhenxi Zhang}[style=chinese]
\ead{zx.zhang3@siat.ac.cn}

\author[1]{Bin Li}[style=chinese, orcid=0000-0002-6508-5071]
\cormark[1]
\ead{b.li2@siat.ac.cn}

\author%
[1]
{Shoujun Zhou}[orcid=0000-0003-3232-6796]
\cormark[1]
\ead{sj.zhou@siat.ac.cn}

\cortext[cor1]{Corresponding author}

\begin{abstract}
Medical image segmentation is crucial for clinical diagnosis and treatment planning, especially when dealing with complex anatomical structures such as vessels. However, accurately segmenting vessels remains challenging due to their small size, intricate edge structures, and susceptibility to artifacts and imaging noise. In this work, we propose VesselSAM, an enhanced version of the Segment Anything Model (SAM), specifically tailored for aortic vessel segmentation.  VesselSAM incorporates AtrousLoRA, a novel module integrating Atrous Attention and Low-Rank Adaptation (LoRA), to enhance segmentation performance. Atrous Attention enables the model to capture multi-scale contextual information, preserving both fine-grained local details and broader global context. LoRA facilitates parameter efficient fine-tuning of the frozen SAM image encoder, reducing the number of trainable parameters and thereby enhancing computational efficiency. We evaluate VesselSAM using two challenging multi-center datasets: the Aortic Vessel Tree (AVT) dataset and the Type-B Aortic Dissection (TBAD) dataset. VesselSAM achieves state-of-the-art performance, attaining DSC scores of 93.50\%, 93.25\%, 93.02\%, and 93.26\% across multi-center datasets. Our results demonstrate  that VesselSAM achieves high segmentation accuracy while utilizing only 7\% of the trainable parameters, significantly reducing computational overhead compared to existing large-scale models. This development paves the way for enhanced AI-based aortic vessel segmentation in clinical environments. The code and models will be released at \url{https://github.com/Adnan-CAS/AtrousLora}.
\end{abstract}



\maketitle
\section{Introduction}

Medical imaging stands at the cutting edge of modern healthcare, serving a vital tool in diagnosing and treating various diseases. Within this domain, medical image segmentation is a critical component aiming to delineate structures such as organs, tumors, and vessels \cite{r1}. Aortic vessel segmentation is crucial for diagnosing cardiovascular diseases, enabling precise vascular health assessments and facilitating interventions such as stent placement and aneurysm monitoring. It plays a crucial role in computer-aided diagnosis, treatment planning, and surgical interventions \cite{r2}. With the rapid advancements in computational resources and the increasing availability of medical data, Vision Transformers (ViTs) have emerged as a revolutionary approach in medical image analysis \cite{r3}. Unlike traditional convolutional models, ViTs employ self-attention mechanisms to capture long-range dependencies and global context, significantly enhancing their ability to model complex structures within medical images \cite{r4, r5}.

\par
This paradigm shift has led to the development of advanced segmentation techniques, such as Segment Anything Model (SAM) \cite{r6}, Swin-Unet \cite{r7}, UNETR \cite{r8}, SAMMed \cite{r9}, and MedSAM \cite{r10}, which leverage the power of ViT’s for highly accurate and computationally efficient segmentation tasks. The SAM enables users to generate segmentation masks through interactive prompts, such as clicks, bounding boxes, and text. Its exceptional zero-shot and few-shot capabilities have demonstrated strong effectiveness in natural image segmentation, garnering significant attention. However, despite the SAM's success in natural image segmentation, recent studies have identified several limitations in its application to medical imaging \cite{r11, r12}. These challenges stem from the inherent differences between natural and medical imaging data. Medical imaging datasets typically exhibit low contrast, ambiguous tissue boundaries, and small regions of interest. These limitations hinder SAM’s ability to generalize effectively without further fine-tuning \cite{r13}.

\par
Recent studies \cite{r14,r15,r16} have sought to fine-tune SAM for medical image segmentation by incorporating domain-specific enhancements. However, fine-tuning these models demands substantial computational resources due to the large number of parameters in foundation models like SAM. Moreover, training large models on limited task-specific data frequently leads to overfitting and suboptimal performance. To address these challenges, parameter-efficient fine-tuning (PEFT) methods, such as Low-Rank Adaptation (LoRA) \cite{r17} have emerged as promising solutions. Several techniques have integrated LoRA into SAM to enhance computational efficiency while preserving performance, particularly in medical image segmentation \cite{r18, r19}. 
\begin{figure*}[t]
	\centering
	\includegraphics[width=0.8\linewidth]{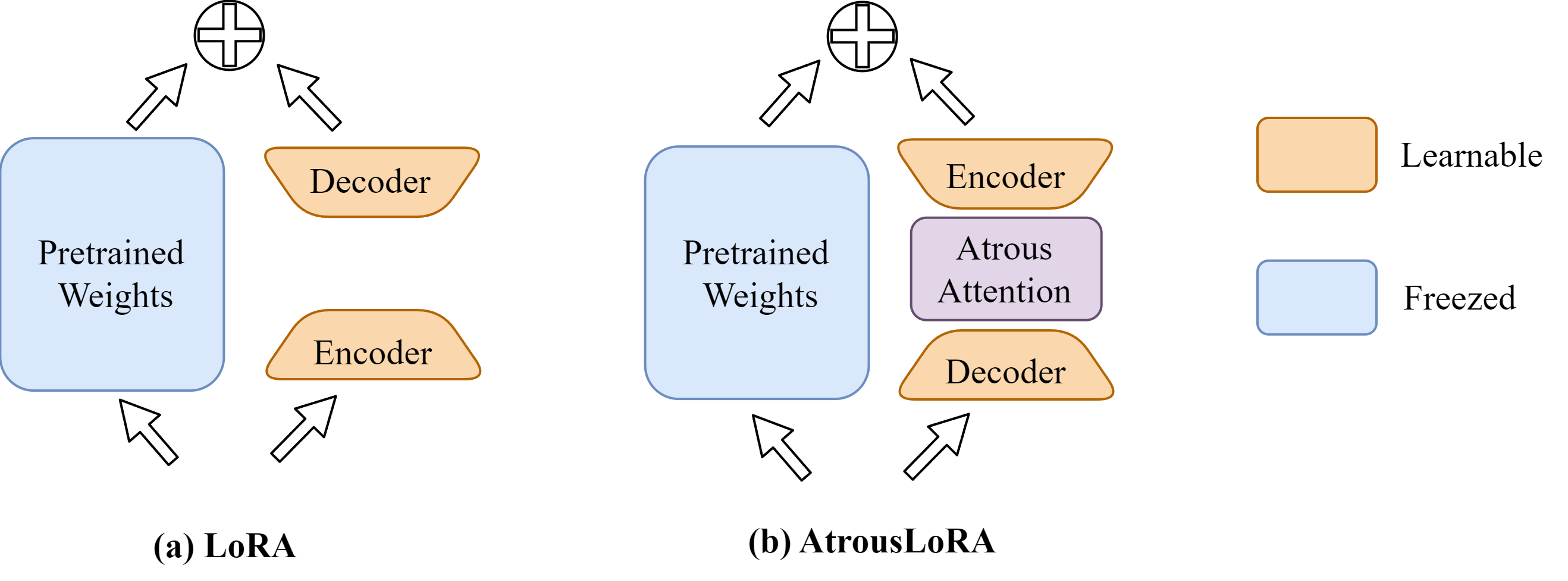}
	\caption{ LoRA and AtrousLoRA: Both LoRA and AtrousLoRA introduce a trainable encoder-decoder structure that operates in parallel with frozen pre-trained weights. (a) LoRA applies a low-rank constraint on the weight updates by factorizing them into smaller matrices (b) AtrousLoRA extends this approach by incorporating Atrous Attention Module into the bottleneck of LoRA, leveraging multi-scale dilated convolution operations for enhanced feature extraction. }
	\label{fig-attention}
\end{figure*}
\par
Despite these advancements, several fundamental intrinsic limitations of SAM persist. The SAM’s image encoder, based on plain ViTs, inherently lacks crucial vision-specific inductive biases needed to capture local patterns and fine details essential for dense predictions in medical imaging \cite{r20}. Additionally, SAM’s ViT-based architecture relies on global attention without integrating regional attention or sparse attention mechanisms, which are vital for focusing on relevant regions and reducing computational overhead \cite{r21}. Although regional attention captures spatial hierarchies at multiple scales, SAM’s reliance on global attention restricts its ability to focus on smaller, intricate regions in medical images. Moreover, the lack of sparse attention inhibits SAM from effectively capturing global context without incurring substantial computational costs. These limitations render SAM susceptible to errors, including the hallucination of small, disconnected components in segmentation \cite{r4, r9}, particularly when modeling structures such as vessels, tumors, or lesions. To enhance the performance of plain ViTs in dense prediction tasks, recent research has combined Transformer and convolutional features \cite{r22, r23}. Recently, the work \cite{r24} integrates atrous attention with ViTs, enabling multi-scale feature extraction while preserving spatial resolution. 

\par Inspired by the work \cite{r24}, we propose VesselSAM, a novel segmentation framework that enhances the SAM by introducing AtrousLoRA, a new adaptation module that integrates Atrous Attention into the Low-Rank Adaptation (LoRA) framework. As illustrated in \cref{fig-attention}, the proposed AtrousLoRA \cref{fig-attention}(b) extends the conventional LoRA design \cref{fig-attention}(a) by embedding an atrous attention block into its bottleneck structure. AtrousLoRA enriches the SAM's encoder with multi-scale contextual reasoning while maintaining the efficiency of LoRA's parameter-reduction approach. At the core of AtrousLoRA lies the Atrous Attention mechanism, which captures both fine-grained and global contextual information by employing dilated attention windows. To further enhance multi-scale representation, Atrous Attention is augmented with Atrous Spatial Pyramid Pooling (ASPP), which utilizes dilated convolutions with varying rates to aggregate features across multiple receptive fields \cite{r22, r23}. This design allows VesselSAM to effectively model vascular structures of varying sizes without compromising spatial resolution. By integrating AtrousLoRA, VesselSAM combines the strengths of global attention and local convolutional inductive biases, making it particularly suitable for complex vascular segmentation tasks. Furthermore, the LoRA-based design ensures parameter-efficient fine-tuning, reducing the need for full model retraining while achieving high segmentation performance across diverse medical imaging scenarios \cite{r18}.
\par
The main contributions of this work are summarized as follows:
\begin{itemize}
    \item  We present VesselSAM, a novel vascular segmentation framework that extends the Segment Anything Model (SAM) through the integration of AtrousLoRA, specifically tailored for aortic vessel segmentation.
    
    \item We propose AtrousLoRA, a module that combines Atrous Attention and Low-Rank Adaptation (LoRA). AtrousLoRA leverages Atrous Spatial Pyramid Pooling (ASPP) to capture multi-scale contextual information and enhances the model's ability to focus on both fine vascular details and broader anatomical structures, while keeping the SAM image encoder frozen.

    \item We introduce a parameter-efficient fine-tuning (PEFT) strategy via AtrousLoRA, enabling VesselSAM to achieve high segmentation accuracy using only 7\% of the trainable parameters, significantly reducing computational cost and data requirements.
     
    \item We evaluate VesselSAM on multiple benchmark datasets, including the Aortic Vessel Tree (AVT) Segmentation dataset and the imageTBAD dataset. Experimental results demonstrate that VesselSAM consistently outperforms existing baseline methods in terms of segmentation accuracy, robustness, and computational efficiency, particularly for aortic vessel segmentation.
    
\end{itemize}

\section{Related Work}

\subsection{ViT and SAM Based Medical Foundation Models}

\par Vision Transformers (ViTs) based medical foundation models have significantly impacted medical image segmentation, with models like UNETR \cite{r8} leading the way. UNETR employs a ViT-based encoder to effectively capture global context while integrating it with a U-Net architecture for precise medical image segmentation. In contrast, SAM-based medical foundation models, which leverage transformer architectures, have exhibited impressive performance across natural image segmentation tasks. However, their direct application to medical image segmentation remains challenging due to unique domain-specific constraints, such as low contrast, complex anatomical structures, and limited labeled data. Recognizing these limitations, MedSAM \cite{r10} sought to enhance SAM’s segmentation performance in the medical domain by freezing the pre-trained image encoder and prompt encoder, while fine-tuning only the lightweight mask decoder on domain-specific medical datasets. This approach effectively leverages SAM’s large-scale pre-trained features while adapting its mask prediction capabilities to medical imaging domain.

\subsection{Parameter-Efficient Model Fine-Tuning}
The concept of Parameter-Efficient Fine-Tuning (PEFT) has emerged as an effective strategy to adapt large foundational models like SAM to specific downstream tasks with minimal additional parameter costs. One prominent PEFT approach, LoRA (Low-Rank Adaptation), has been successfully incorporated into SAM-based models. For instance, SAMed \cite{r11} applied LoRA to SAM’s frozen image encoder, fine-tuning the LoRA layers, the prompt encoder, and the mask decoder together on medical datasets like Synapse multiorgan, demonstrating significant performance improvements. Similarly, SAMAdp \cite{r19} introduced a lightweight adapter module to enhance SAM's segmentation performance in challenging tasks. By integrating task-specific prompts and adapters, SAMAdp improves segmentation accuracy while maintaining computational efficiency, demonstrating broad adaptability across diverse domains. Other works have pursued different approaches to optimize SAM for medical imaging applications. SAMMed \cite{r9} systematically evaluated SAM across 53 public medical imaging datasets, revealing that while SAM demonstrates strong zero-shot segmentation capabilities, its performance often degrades without fine-tuning, reinforcing the need for domain-specific adaptation.

\subsection{Atrous Convolution in ViTs}
Atrous Convolution (dilated convolution) has emerged as a powerful technique in Vision Transformers (ViTs) to enhance both local feature extraction and global contextual modeling, which are critical for segmentation tasks \cite{Liu2022,TONG2024213,Lam2021}. Atrous convolution expands the receptive field by introducing pixel “skipping”, enabling the model to capture multi-scale spatial dependencies without downsampling. This preserves fine-grained details while improving the ability to model broader spatial relationships. Initially introduced in DeepLab \cite{r20} for convolutional networks, Atrous Convolution has proven highly effective in extracting multi-scale features, which is crucial for handling segmentation tasks involving objects of varying sizes. In ViTs, where image features are typically processed as non-overlapping patches, integrating Atrous Convolutions enhances the model’s ability to learn hierarchical spatial dependencies. Specifically, Atrous Spatial Pyramid Pooling (ASPP) modules apply dilated convolutions at multiple rates, allowing the model to capture multi-scale contextual information \cite{Yu2015}, bridging the gap between local interactions and global dependencies. This approach is particularly beneficial in tasks requiring detailed segmentation, where capturing both local fine details and global context is necessary for accurate predictions. Recent advancements have shown that Atrous Convolutions are crucial for improving the performance of ViTs in segmentation tasks, particularly in domains such as medical imaging. In our model, we leverage the power of ASPP and Attention mechanisms to enhance the ViT encoder’s ability to capture both local priors and global context, effectively enabling the model to handle complex, high-resolution segmentation tasks with greater accuracy.
\begin{figure*}[t]
	\centering
	   \includegraphics[width=\linewidth]{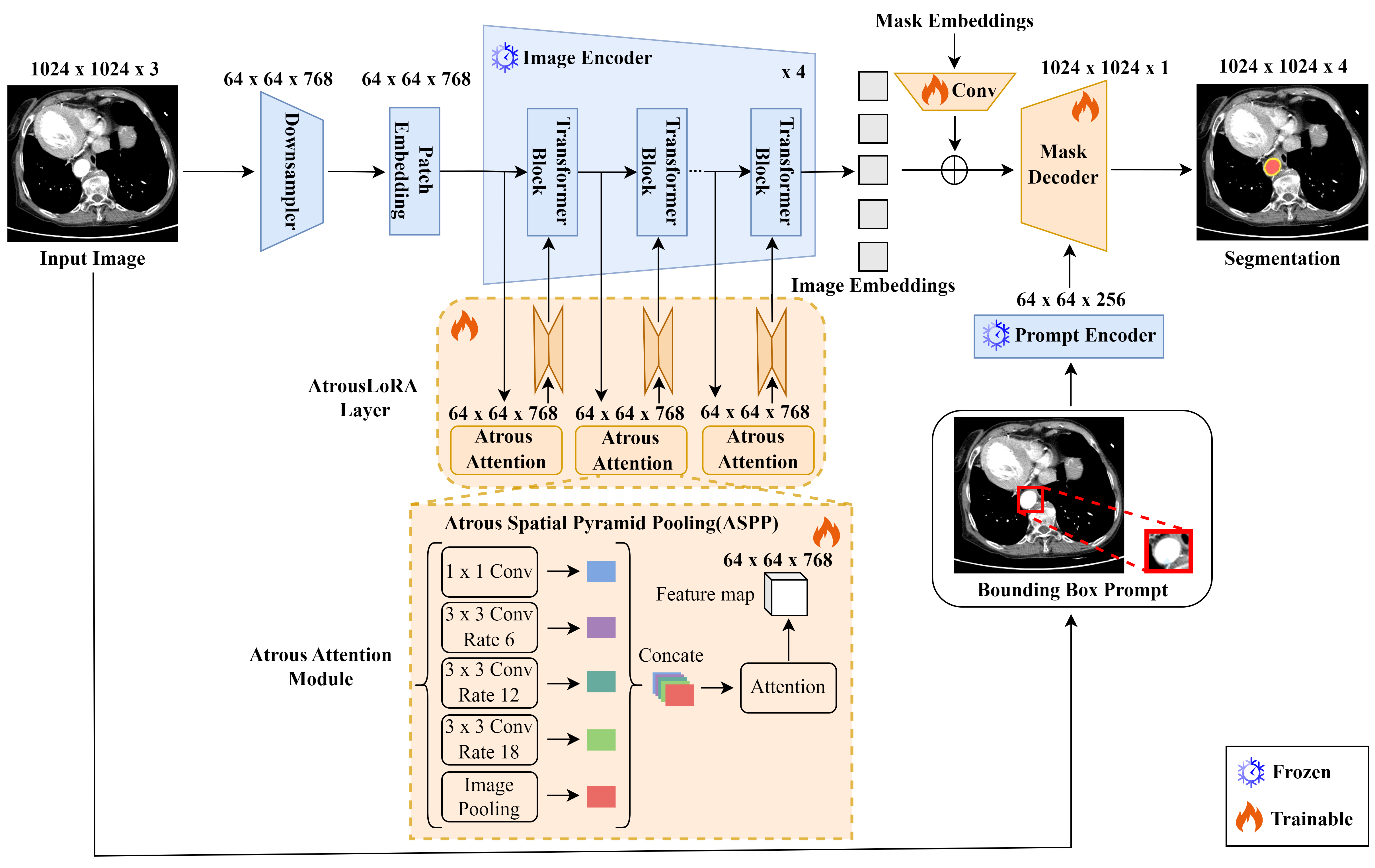}
	\caption{The architecture of the proposed VesselSAM framework. The model combines frozen pre-trained transformer blocks with learnable AtrousLoRA layers, enabling enhanced multi-scale feature extraction through Atrous Attention Module. Image embeddings generated by the frozen image encoder are fused with a bounding box prompt from the frozen prompt encoder and processed by the learnable mask decoder to produce the final segmentation. Blue represents frozen components, while orange denotes learnable parameters.}
	\label{fig1}
\end{figure*}

\section{Methodology}
\subsection{Overview}
VesselSAM is a promptable segmentation model designed to enhance vascular structure segmentation in medical imaging. It builds upon the Segment Anything Model (SAM) framework while integrating AtrousLoRA, a novel module that combines Atrous Attention with Low-Rank Adaptation (LoRA) to improve segmentation accuracy and computational efficiency. To preserve the rich pre-trained representations of SAM, both the image encoder and prompt encoder remain frozen, while Atrous Attention and LoRA layers enhance the model’s ability to capture multi-scale contextual information and optimize training efficiency. The Atrous Attention module expands the receptive field through dilated convolutions, enabling the segmentation of both fine-grained vascular structures and broader anatomical features without increasing computational cost. Meanwhile, AtrousLoRA layers inserted within the frozen image encoder apply low-rank projections to reduce the number of trainable parameters, allowing for efficient fine-tuning while maintaining the integrity of the pre-trained backbone. The final segmentation output is generated by the mask decoder, which refines the fused embeddings from the image and prompt encoders through cross-attention mechanisms, ensuring accurate and robust segmentation performance. By leveraging AtrousLoRA, VesselSAM achieves state-of-the-art segmentation accuracy while significantly reducing computational overhead, making it well-suited for medical image segmentation tasks, particularly vascular segmentation in aortic imaging.

\subsection{Preliminary: SAM architecture}
The SAM \cite{r6} is a prompt-based segmentation framework composed of three main components: the Image Encoder, Prompt Encoder, and Mask Decoder. The Image Encoder is based on a ViT, which processes input images using 16 × 16 pixel patches through transformer blocks to capture image features, resulting in image embedding. The Prompt Encoder handles various prompts, including points, bounding boxes and masks, converting them into feature vectors that guide the segmentation. These prompt embeddings enable SAM to focus on specific regions of interest within an image, improving segmentation accuracy and adaptability. The Mask Decoder is a two-layer transformer-based module that fuses image embedding and prompt features using cross-attention mechanisms. To refine feature representations and ensure precise mask generation, the decoder incorporates a Multi-Layer Perceptron (MLP) for feature refinement and dimensionality alignment. Additionally, convolutional layers are utilized for upsampling, allowing the model to produce high-resolution segmentation masks.

\subsection{VesselSAM}
The VesselSAM architecture builds on the foundation of SAM framework, incorporating key modifications to improve aortic vessel segmentation. As illustrated in \cref{fig1}, VesselSAM integrates the Atrous Attention module and LoRA layers, designed to capture multi-scale features and reduce the number of trainable parameters while maintaining segmentation accuracy.

In this design, the image encoder and prompt encoder from the original SAM architecture are frozen to retain their powerful pre-trained features. The image encoder, based on a Vision Transformer (ViT), extracts rich visual features from the input medical images. The prompt encoder processes sparse prompts such as points or bounding boxes, which guide the segmentation process by focusing on specific regions of interest in the image.

To enhance the model’s ability to capture both local and global features, the Atrous Attention module is integrated into the frozen image encoder. This module utilizes dilated convolutions to expand the receptive field, allowing the model to capture multi-scale features, which are crucial for medical images like small tumors or vascular boundaries.

Additionally, LoRA (Low-Rank Adaptation) layers are inserted between the transformer blocks in the image encoder. These layers compress the transformer features into a low-rank space and then re-project them, allowing efficient adaptation of the features while preserving the frozen transformer parameters. This modification improves training efficiency, reducing the number of trainable parameters and enhancing the model’s performance with fewer resources.

The final segmentation is generated by the mask decoder, which consists of a lightweight transformer decoder and a segmentation head. During training, the mask decoder is fine-tuned to refine the fused embeddings from the image and prompt encoders using cross-attention mechanisms. This ensures that the model is able to accurately segment fine-grained details, such as vascular structures, while also preserving broader anatomical context.

\subsection{LoRA and AtrousLoRA}

LoRA \cite{r17} has emerged as a PEFT method, enabling task-specific adaptations of pre-trained models while significantly reducing computational and memory overhead. LoRA introduces low-rank trainable matrices to approximate weight updates, effectively bypassing the need to fine-tune the entire model (\cref{fig-attention}(a)). Specifically, it adds two small matrices, \( W_b \) and \( W_a \), while keeping the original weights \( W_O \) frozen during training. Given a pre-trained weight matrix \( W_O \in \mathbb{R}^{C_{\text{out}} \times C_{\text{in}}} \), LoRA modifies the forward pass of the model as 
\begin{equation}
y = W_O x + W_b W_a x  \label{eq:LoraEq}
\end{equation} 
where \( W_O \) is the frozen pre-trained weight matrix, \( W_b \in \mathbb{R}^{r \times C_{\text{in}}} \) and \( W_a \in \mathbb{R}^{C_{\text{out}} \times r} \) are the low-rank encoder and decoder matrices, and \( r \) is the rank of the decomposition, with \( r \ll \min(C_{\text{in}}, C_{\text{out}}) \). Here, \( x \in \mathbb{R}^{B \times C_{\text{in}}} \) represents the input, where \( B \) is the batch size. 
\par While LoRA is highly efficient for adapting pre-trained models, it lacks the ability to explicitly capture multi-scale contextual information, which is critical for vision tasks such as image segmentation and dense prediction. To address this limitation, we introduced AtrousLoRA which incorporates atrous (dilated) convolutions into the LoRA framework (\cref{fig-attention}(b)). Atrous convolutions expand the receptive field of the model without increasing the number of parameters, enabling it to capture both local and global dependencies. 

Mathematically, with AtrousLoRA Eq.(\ref{eq:LoraEq}) changes to:
\begin{equation}
y = W_O x + W_b \cdot \text{AtrousAttention}(W_a x) \label{eq:ATLoraEq}
\end{equation} where \( W_O \in \mathbb{R}^{C_{\text{out}} \times C_{\text{in}}} \) is the frozen pre-trained weight matrix, \( W_a \in \mathbb{R}^{r \times C_{\text{in}}} \) and \( W_b \in \mathbb{R}^{C_{\text{out}} \times r} \) are the low-rank encoder and decoder matrices, and \( x \in \mathbb{R}^{B \times C_{\text{in}} \times H \times W} \) is the input feature map. In this case, \( B \) represents the batch size, \( C_{\text{in}} \) and \( C_{\text{out}} \) are the input and output channels, and \( H \) and \( W \) represent the height and width of the feature maps. The AtrousAttention module applies atrous convolutions with predefined dilation rates to \( W_a x \), effectively capturing multi-scale contextual features.  The AtrousAttention can be formulated as:
\begin{equation}
\text{AtrousAttention}(W_a x) = Y_{\text{ASPP}} \odot A_{\text{sigmoid}} \label{eq:AtrousAttentionEq}
\end{equation} where \( Y_{\text{ASPP}} \) is the output of the Atrous Spatial Pyramid Pooling (ASPP) module and \( A_{\text{sigmoid}} \) is the attention map generated by the attention mechanism. The element-wise product \( Y_{\text{ASPP}} \odot A_{\text{sigmoid}} \) combines the multi-scale features from the ASPP module with the attention map, resulting in an attention-weighted feature map that enhances important regions and suppresses less relevant ones. This mechanism enables AtrousLoRA to focus on the most salient features while maintaining contextual information across multiple scales.

\subsection{Atrous Spatial Pyramid Pooling}
\begin{figure*}[t]
	\centering
	   \includegraphics[width=0.7\linewidth]{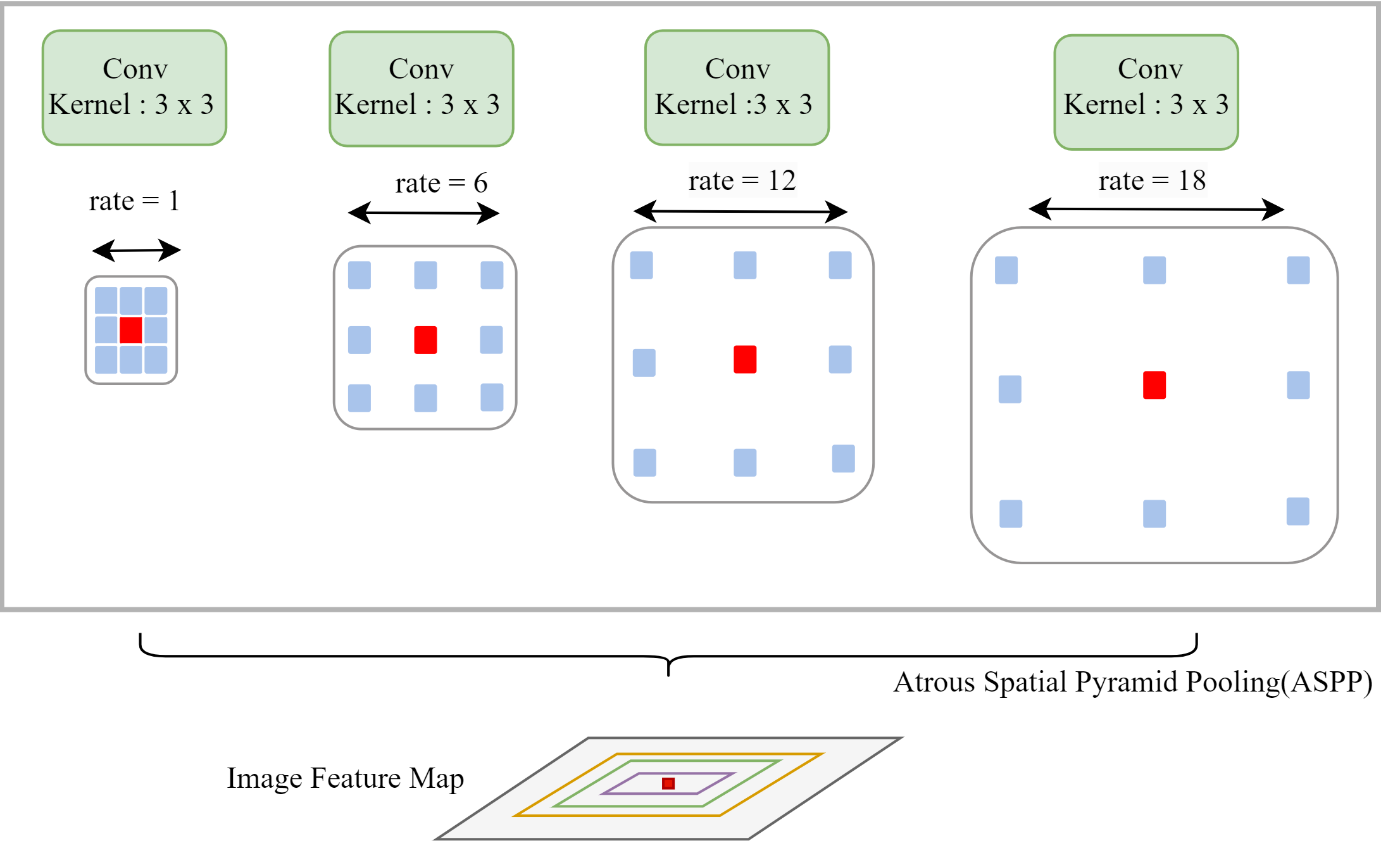}
	\caption{Atrous Spatial Pyramid Pooling (ASPP) captures multi-scale features by applying parallel convolutional filters with different dilation rates. Each configuration results in a distinct effective receptive field, illustrated in different colors to highlight the varying field-of-views.}
	\label{fig-ASPP}
\end{figure*}

Atrous Spatial Pyramid Pooling (ASPP), originally proposed by \cite{r21}, capable of capturing multi-scale contextual information, as illustrated in \cref{fig-ASPP}. In our work, ASPP is integrated into the Atrous Attention Module to enhance the segmentation of vascular structures in medical images. By leveraging dilated convolutions with varying dilation rates, ASPP enables the model to capture both fine details and broader contextual information without sacrificing resolution. This capability is particularly important for accurately segmenting blood vessels, as it allows the model to understand both local features and their spatial relationships within the image.

Mathematically, ASPP operates by applying dilated convolutions with multiple dilation rates \( d_i \in \{d_1, d_2, \dots, d_n\} \), where each rate \( d_i \) extracts features at a specific scale. For a given input feature map \( X \in \mathbb{R}^{B \times C \times H \times W} \), the dilated convolution operation for each rate \( d_i \) is defined as.
\begin{equation}
Y_i = f_{\text{dil}}(X; W_i, d_i) = X *_{d_i} W_i \label{eq:ASPPEq1}
\end{equation} where \( *_{d_i} \) denotes the dilated convolution operation, and \( W_i \) represents the convolutional filter with dilation rate \( d_i \).

In addition to the multi-scale dilated convolutions, ASPP incorporates a global average pooling (GAP) operation to capture the global context of the input feature map. Mathematically the GAP operation is defined as:
\begin{equation}
Z = \frac{1}{H \times W} \sum_{h=1}^H \sum_{w=1}^W X_{b,c,h,w} \label{eq:ASPPEq2}
\end{equation} where \( Z \in \mathbb{R}^{B \times C \times 1 \times 1} \) represents the globally pooled feature map, summarizing the spatial information into a single vector per channel. The outputs of the dilated convolutions \( Y_i \) and the global average pooling \( Z \) are then concatenated into a single feature map as expressed in Eq. \ref{eq:ASPPEq3}. 
\begin{equation}
Y_{\text{concat}} = [Y_1, Y_2, \dots, Y_n, Z] \label{eq:ASPPEq3}
\end{equation} where \( Y_{\text{concat}} \in \mathbb{R}^{B \times C' \times H \times W} \) combines multi-scale features, enabling the model to capture both local and global contextual information. To reduce the dimensionality of the concatenated feature map, a \( 1 \times 1 \) convolution is applied.
\begin{equation}
Y_{\text{ASPP}} = f_{\text{1x1}}(Y_{\text{concat}}) = W_{\text{1x1}} \cdot Y_{\text{concat}} + b_{\text{1x1}} \label{eq:ASPPEq4}
\end{equation} where \( W_{\text{1x1}} \) and \( b_{\text{1x1}} \) are the weight and bias of the \( 1 \times 1 \) convolution, respectively. Finally, a non-linear activation function ReLU \cite{agarap2018deep} is applied.
\begin{equation}
Y_{\text{ASPP}} = \text{ReLU}(Y_{\text{ASPP}}) \label{eq:ASPPEq5}
\end{equation}

\begin{figure*}[ht]
	\centering
	\includegraphics[width=0.99\linewidth]{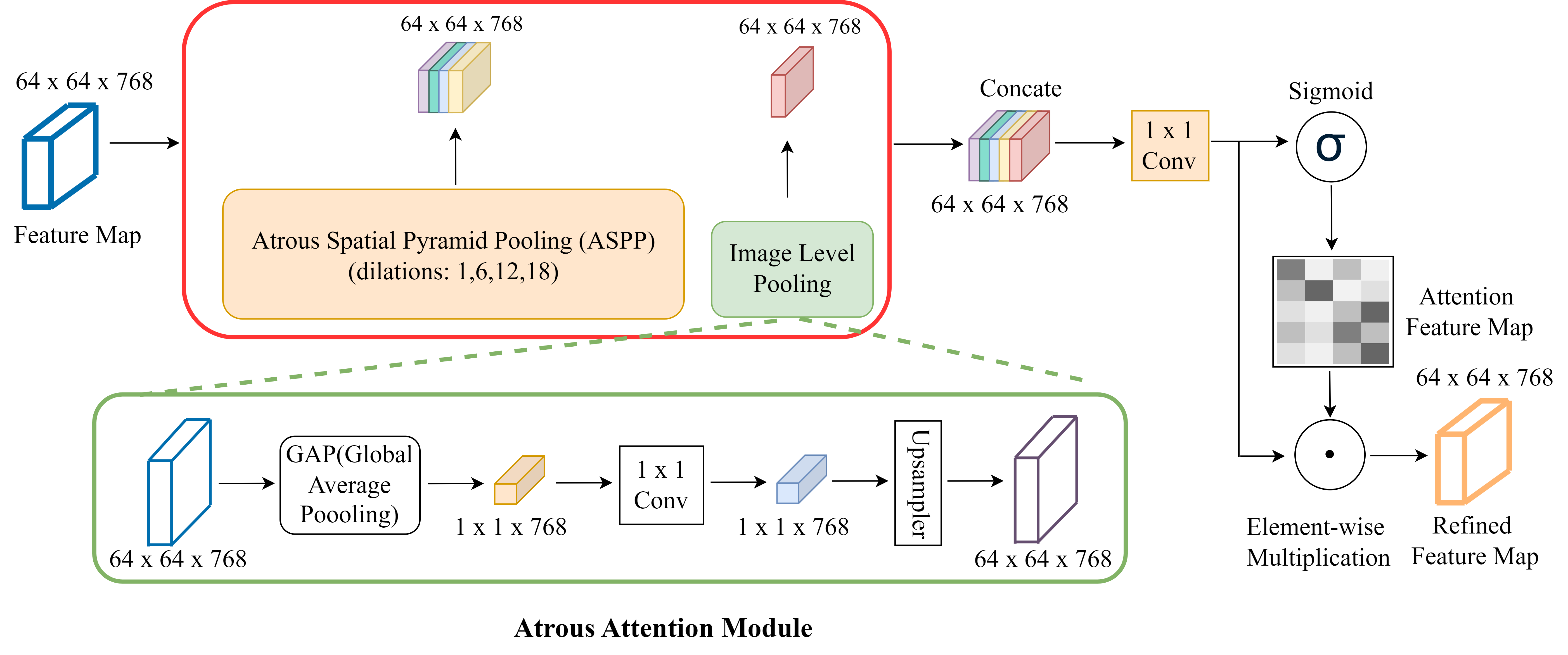}
	\caption{Atrous Attention Module: The Atrous Attention Module features an Atrous Spatial Pyramid Pooling (ASPP) module with various dilation rates, global image-level pooling, and an attention mechanism that refines feature maps through element-wise multiplication with attention weights.}
	\label{fig2}
\end{figure*}

\subsection{Atrous Attention Module}
The Atrous Attention Module is introduced as a novel attention mechanism for Vision Transformers, designed to fuse regional and sparse attention effectively, as shown in \cref{fig2}. This approach allows us to capture both global context and local detail with efficient computational complexity, while preserving the hierarchical information present in medical images. Inspired by atrous convolution \cite{r24}, which expands the receptive field by skipping rows and columns in the input feature map without increasing the number of parameters. Atrous Attention enables VesselSAM to focus on relevant anatomical structures across multiple scales. The process is shown in Algorithm \ref{alg1}.
\par The data flow within the Atrous Attention Module starts by passing the input feature map $ X \in \mathbb{R}^{B \times C \times H \times W} $
through the ASPP, which applies dilated convolutions at different rates \( d_i \) to capture features at various scales. Each atrous convolution produces an output feature map $ Y_i = f(X; W_i, d_i) $ where \( W_i \) are the convolution weights and \( d_i \) is the dilation rate. 

\begin{algorithm}
\caption{Atrous Attention Module}
\begin{normalsize}
\begin{algorithmic}[1]
\STATE \textbf{Input:} \( X \in \mathbb{R}^{B \times C \times H \times W} \) 

\STATE \textbf{Output:} \( Y_{\text{out}} \in \mathbb{R}^{B \times C \times H' \times W'} \) 

\STATE  \( Y_i = f_{\text{dil}}(X; W_i, d_i), \quad i = 1 \dots n \) \\
\STATE  \( Z = \frac{1}{H W} \sum_{h=1}^{H} \sum_{w=1}^{W} X_{b,c,h,w} \) \\
\STATE  \( Y = [Y_1, Y_2, \dots, Y_n, Z] \) \\
\STATE  \( Y_{\text{ASPP}} = \text{ReLU}(\text{BN}(f_{\text{1x1}}(Y))) \) \\
\STATE  \( A = \sigma(f_{\text{1x1}}(Y_{\text{ASPP}})) \) \\
\STATE  \( Y_{\text{out}} = Y_{\text{ASPP}} \odot A \) \\

\end{algorithmic}
\end{normalsize}
\label{alg1}
\end{algorithm}

Additionally, global average pooling is applied to the input
\( X \) to obtain $Z = f_{\text{1x1}}(\text{GAP}(X))$. The outputs from ASPP, including the atrous convolutions \( Y_i \) and the global pooling result \( Z \), are concatenated into a Concatenated Feature Map $ Y = [Y_1, Y_2, \dots, Y_n, Z] $. This concatenated feature map then goes through a 1x1 Convolution, reducing it to the desired number of output channels, followed by Batch Normalization (BN) \cite{ioffe2015batch} and ReLU activation, generating the ASPP Output$ Y_{\text{ASPP}} = \text{ReLU}(\text{BN}(f_{\text{1x1}}(Y)))$. 

\par This output is further processed through another 1x1 Convolution to create the Attention Map $ A = f_{\text{1x1}}(Y_{\text{ASPP}}) $  where \( A \in \mathbb{R}^{B \times 1 \times H \times W} \)  is the attention map and a Sigmoid activation is applied to obtain $ A_{\text{sigmoid}} = \sigma(A) $ which constrains the attention values between 0 and 1 that is where \( A_{\text{sigmoid}} \in [0, 1]^{B \times 1 \times H \times W} \). Finally, the ASPP Output is multiplied element-wise with the Attention Map, producing the Final Output $ Y_{\text{out}} = Y_{\text{ASPP}} \odot A_{\text{sigmoid}}, $ The result is an attention-weighted feature map \( Y_{\text{out}} \in \mathbb{R}^{B \times C' \times H \times W} \), where important regions of the feature map are enhanced, and less important regions are suppressed. This mechanism enhances VesselSAM’s ability to focus on the most important features, improving segmentation accuracy while maintaining context from multiple scales.

\subsection{Prompt Encoder And Mask Decoder}
In VesselSAM, the Prompt Encoder remains frozen, ensuring the stability of the pre-trained parameters while allowing for efficient processing of user prompts. In our case, the prompts are provided in the form of bounding boxes, which are represented by their top-left and bottom-right corner points. Each corner point is mapped into a 256-dimensional embedding, which serves as the input to the segmentation process. By freezing the prompt encoder, VesselSAM enables real-time interaction. The image embedding can be precomputed, allowing users to dynamically provide bounding-box input without the need for retraining.

On the other hand, the Mask Decoder in VesselSAM is fully trainable and plays a crucial role in producing the segmentation output. The decoder architecture includes two transformer layers, which are responsible for fusing the image embedding with the prompt embeddings through cross-attention. This fusion allows the bounding box information to guide the segmentation task effectively. The mask decoder employs two transposed convolution layers to upsample the combined embedding to a resolution of 256 × 256 while ensuring a high level of detail is retained in the final segmentation mask. The output is then passed through a sigmoid activation function followed by bi-linear interpolation to match the resolution of the original input image thereby producing the final high-resolution mask.

\subsection{Loss Function and Evaluation Metrics}
We have used a combined loss function comprising an unweighted sum between cross-entropy loss and Dice loss, which has been widely adopted for its robustness in medical image segmentation tasks \cite{r10}. As detailed in Eq.\eqref{eq:LossEq1}, Eq.\eqref{eq:LossEq2}, and Eq.\eqref{eq:LossEq3}, where \( P \) represents the predicted segmentation output and \( T \) denote the corresponding ground truth. For each voxel \( j \), \( p_j \) and \( t_j \) correspond to the predicted and ground truth values, respectively. The total number of voxels in the image is denoted by \( M \). The binary cross-entropy loss is defined as:

\begin{equation}
L_{CE} = -\frac{1}{M}\sum_{j=1}^{M}[t_j \log p_j + (1 - t_j)\log(1 - p_j)], \label{eq:LossEq1}
\end{equation} where \( L_{CE} \) quantifies the pixel-wise classification accuracy. The Dice loss, which measures the overlap between the predicted and ground truth regions, is given by:

\begin{equation}
L_{D} = 1 - \frac{2 \sum_{j=1}^{M} t_j p_j}{\sum_{j=1}^{M} (t_j)^2 + \sum_{j=1}^{M} (p_j)^2}, \label{eq:LossEq2}
\end{equation}

The final loss \( L \) is computed as the sum of the Dice loss and the cross-entropy loss :

\begin{equation}
L = L_{Dice} + L_{CE}  \label{eq:LossEq3}
\end{equation}

This combined loss function ensures effective training by balancing region-based overlap and pixel-wise classification accuracy, making it suitable for a wide range of medical image segmentation tasks.

To evaluate the performance of the segmentation model, we employed two metrics: Dice Similarity Coefficient (DSC) and Hausdorff Distance (HD). The DSC measures the spatial overlap between the predicted segmentation \( P \) and the ground truth \( T \), and is defined as:

\begin{equation}
DSC(P, T) = \frac{2|P \cap T|}{|P| + |T|} \label{eq:DiceEq1}
\end{equation} where \( |P \cap T| \) represents the intersection of the predicted and ground truth regions, and \( |P| \) and \( |T| \) denote the sizes of the predicted and ground truth regions, respectively. A higher DSC value indicates better segmentation accuracy, with a maximum value of 1 indicating perfect overlap.

The Hausdorff Distance (HD) quantifies the maximum distance between the boundaries of the predicted segmentation and the ground truth. It is defined as:

\begin{equation}
HD(P, T) = \max \left( \sup_{x \in \partial P} \inf_{y \in \partial T} d(x, y), \sup_{y \in \partial T} \inf_{x \in \partial P} d(x, y) \right) \label{eq:HDEq1}
\end{equation} where \( \partial P \) and \( \partial T \) represent the boundary points of the predicted and ground truth regions respectively and \( d(x, y) \) is the Euclidean distance between points \( x \) and \( y \). A lower HD value indicates better boundary alignment between the predicted and ground truth segmentations.

\section{Experiments}
\subsection{Datasets}
In our experiments, we utilized two key datasets to evaluate the effectiveness of the proposed VesselSAM model in complex medical segmentation tasks. The Aortic Vessel Tree (AVT) Segmentation dataset \cite{r25} comprises 56 contrast-enhanced CT angiography (CTA) scans collected from three sources: the KiTS Grand Challenge, the Rider Lung CT dataset, and Dongyang Hospital. Among these, 38 cases were designated for training, while the remaining 18 were used for testing. All slices were resampled to a spatial resolution of 1 mm × 1 mm, with Hounsfield Unit (HU) values normalized to [0, 1]. Additionally, the TBAD dataset \cite{r26}, comprising 100 CTA images from Guangdong Provincial People's Hospital, was utilized for segmenting True Lumen (TL), False Lumen (FL), and False Lumen Thrombus (FLT) in Type-B Aortic Dissection (TBAD) cases. To conform to the SAM requirements, both the AVT and TBAD datasets were converted from 3D CTA volumes into 2D slices. Each 3D scan was converted into NumPy arrays, with all slices resampled to a uniform resolution of 1 mm × 1 mm. Voxel intensity values were normalized using standard CT window settings [400, 40]. Ground truth masks were refined by removing labels of irrelevant structures and small objects, using thresholds of 1000 voxels for 3D volumes and 100 pixels for individual 2D slices. Only non-zero slices were retained, and intensity normalization was applied. The processed 2D slices were then resized to 1024 × 1024 pixels and converted into three-channel images by duplicating the grayscale slice across three channels (1024 × 1024 × 3), ensuring compatibility with SAM’s input format.

\subsection{Implementation Details}
All experiments were conducted using the VesselSAM model implemented with the PyTorch deep learning library. VesselSAM is based on the SAM architecture, employing a ViT-Base image encoder initialized with pre-trained weights. During training, the parameters of the image encoder and prompt encoder remained frozen, while fine-tuning was applied exclusively to the mask decoder and the integrated AtrousLoRA modules. The AtrousLoRA module comprises ASPP and attention mechanisms. Specifically, the ASPP utilized dilated convolutions with dilation rates of 1, 6, 12, and 18, enabling the capture of multi-scale contextual information critical for accurate segmentation. Additionally, LoRA layers with a low-rank dimension of 4 were integrated to achieve an optimal trade-off between model accuracy and computational efficiency by reducing the number of trainable parameters to approximately 7\% of the total parameters.

The model was optimized using the AdamW optimizer with an initial learning rate of 1e-4 and a weight decay of 0.01. Training proceeded for a total of 100 epochs with a batch size of 8. To further improve computational efficiency, mixed-precision training was employed. Data augmentation techniques included random perturbations to bounding box coordinates to improve the model’s generalizability. All experiments were performed on a single NVIDIA H100 GPU with 80GB VRAM. The model was evaluated on two challenging benchmark datasets: the Aortic Vessel Tree (AVT) and the Type-B Aortic Dissection (TBAD). Performance metrics included the Dice Similarity Coefficient (DSC) and the Hausdorff Distance (HD).

\begin{table*}[t]
\caption{Performance Comparison of our Proposed VesselSAM with other ViTBased and SAMbased Models on AVT dataset}
\label{tab1}
\centering
\resizebox{\textwidth}{!}{%

\begin{tabular}{llcccccc}
\toprule
\multicolumn{8}{c}{\textbf{Big Model}} \\ 
\toprule
\multirow{2}{*}{\centering Methods} & \multirow{2}{*}{\#Params (M) /Ratio (\%)}  &  
\multicolumn{2}{c}{AVT-Dataset Dongyang Hospital} & 
\multicolumn{2}{c}{AVT-Dataset Rider Hospital} & 
\multicolumn{2}{c}{AVT-Dataset KiTs Hospital} \\ 
\cline{3-8}
 &  &  DSC(\%) \(\uparrow\) & HD(mm) \(\downarrow\) & DSC(\%) \(\uparrow\) & HD(mm) \(\downarrow\) & DSC(\%) \(\uparrow\) & HD(mm) \(\downarrow\) \\ 
\midrule
UNETR \cite{r8} & 92.5 / 100 & 89.38 & 4.15 & 88.04 & 4.39 & 88.69 & 4.28 \\ 
SAM-ViTb \cite{r6} & 91.0 / 100 & 81.12 & 9.85 & 79.93 & 10.20 & 80.50 & 10.00 \\ 
MedGIFT \cite{r28} & 120.7 / 100 & 88.70 & 4.27 & 87.50 & 4.41 & 87.09 & 4.37 \\ 
MedSAM-Vanilla \cite{r10} & 93.7 / 100 & 89.50 & 4.13 & 87.04 & 4.46 & 88.65 & 4.27 \\ 
MedSAM-FT \cite{r10} & 93.7 / 100 & \underline{92.49} & \underline{3.64} & \underline{90.35} & \underline{4.01} & \underline{91.45} & \underline{3.95} \\ 
SAMMed-Vanilla \cite{r9} & 91.0 / 100 & 88.02 & 4.37 & 87.30 & 4.43 & 87.20 & 4.45 \\ 
SAMMed-FT \cite{r9} & 91.0 / 100 & 89.76 & 4.10 & 88.25 & 4.32 & 88.75 & 4.28 \\ 
\midrule
\multicolumn{8}{c}{\textbf{Small Model}} \\ 
\midrule
SAMed-FT \cite{r11} & 6.3 / 6.7 & 88.23 & 4.34 & 89.45 & 4.14 & 88.80 & 4.24 \\ 
SAMAdp-FT \cite{r19} & 4.1 / 4.3 & 90.30 & 4.02 & 89.75 & 4.10 & 89.90 & 4.05 \\ 
\midrule
\textbf{VesselSAM} & \textbf{6.8 / 7.2} &  \textbf{93.50} & \textbf{3.56} &  \textbf{93.25} & \textbf{3.59} & \textbf{93.02} & \textbf{3.64} \\ 
\bottomrule
\end{tabular}%
}
\vspace*{0.02cm} 
    \newline
    \small
    \justifying
    Note: Bold indicates the best results and underline denotes the second best results. "Vanilla" refers to versions using pre-trained weights, while "FT" indicates fine-tuned versions on AVT datasets. 
\end{table*}

\begin{table}[ht]

    \caption{Performance Comparison of our Proposed VesselSAM with other ViTBased and SAMbased Models on TBAD Dataset} \label{tab2}
    \hfill
    \begin{minipage}{1.0\textwidth}
    \centering
    \scriptsize 
    \begin{tabular} {lccc}
        \toprule
        \multicolumn{4}{c}{\textbf{Big Model}} \\ 
        \midrule
        Methods & \#Parms(M) / Ratio(\%) & DSC(\%) \(\uparrow\) & HD(mm) \(\downarrow\) \\
        
        \midrule
        UNETR \cite{r8} & 92.5 / 100 & 89.20 & 4.18 \\
        SAM-VitB \cite{r6} & 91.0 / 100 & 79.53 & 10.15 \\
        MedGIFT \cite{r28} & 120.7 / 100 & 87.60 & 4.49 \\
        MedSAM-Vanilla \cite{r10} & 93.7 / 100 & 88.40 & 4.29 \\
        MedSAM-FT \cite{r10} & 93.7 / 100 & \underline{92.20} & \underline{3.63} \\
        SAMMed-Vanilla \cite{r9} & 91.0 / 100 & 89.40 & 4.14 \\
        SAMMed-FT \cite{r9} & 91.0 / 100 & 87.40 & 4.40 \\
        \midrule
        \multicolumn{4}{c}{\textbf{Small Model}} \\ 
        \midrule
        SAMed-FT \cite{r11} & 6.3 / 6.7 & 88.20 & 4.43 \\
        SAMAdp-FT \cite{r19} & 4.1 / 4.3 & 89.71 & 4.14 \\
        \midrule
        \textbf{VesselSAM} & \textbf{6.8 / 7.2} & \textbf{93.26} & \textbf{3.58} \\
        \bottomrule
    \end{tabular}
    \end{minipage}
    \hfill
\vspace{0.02cm}
    \newline
    \small
    \justifying
    Note: Bold indicates the best results and underline denotes the second best results. "Vanilla" refers to versions using pre-trained weights, while "FT" indicates fine-tuned versions on TBAD dataset.
\end{table}
\begin{table*}[H]

 \caption{Performance Comparison of our Proposed VesselSAM with other Non-SAM methods on AVT and TBAD Dataset} \label{tab3}
\hfill
\begin{minipage}{1.0\textwidth}
\centering
\scriptsize 

\begin{tabular}{llccccc}
\multirow{2}{*}{\centering Methods}  &  
\multirow{2}{*}{\centering Params Size (M)}  & 
\multicolumn{2}{c}{AVT \cite{r19}}  & 
\multicolumn{2}{c}{TBAD \cite{r19}} \\ 
\cline{3-6}
 &&   DSC(\%) \(\uparrow\) & HD(mm) \(\downarrow\) & DSC(\%) \(\uparrow\) & HD(mm) \(\downarrow\) \\ 
\midrule

Xiong \cite{r35} & 75.4 & 88.60 & 7.72 & 90.70 & 5.51  \\ 
Sieren \cite{r36} &  47.2 & 87.90 & 10.01 & 89.40 & 5.22  \\ 
Li \cite{r37} &  51.6 & 88.50 & 7.43 & 91.80 & 3.41  \\ 
Zhao \cite{r38} & 40.1 & 88.10 & 9.45 & 90.30 & 5.43  \\ 
Feiger \cite{r39} &  63.4 & 88.30 & 8.76 & 90.20 & 5.02  \\ 
Wobben \cite{r40} & 55.8 & 88.70 & 7.54 & 90.60 & 4.57  \\ 
Song \cite{r41} & 56.4 & 88.40 & 7.86 & 90.70 & 6.06  \\ 
Hahn \cite{r42} & 33.7 & 88.00 & 12.35 & 88.50 & 8.65  \\ 
Abdolmanafi \cite{r43}  & 53.4 & 88.20 & 9.12 & 90.30 & 5.77  \\ 
Chen \cite{r44} & 52.2 & 88.90 & 8.36 & 90.50 & 5.63  \\ 
Lyu \cite{r45} & 57.5 &  88.80 & 8.41 &  90.20 & 6.53  \\
Zhao \cite{r46}  & 46.8 & 89.80 & 7.13 & 91.20 & 4.02  \\ 
Deng \cite{r47} & 40.3 & 88.30 & 10.36 & 89.40 & 5.88  \\ 
Yu \cite{r48} & 34.6 &  88.20 & 9.33 &  89.70 & 6.26  \\ 
Cheng \cite{r49} & 88.20 & 88.00 & 9.88 & 90.10 & 4.97  \\ 
Cao \cite{r50} &  42.1 &  88.10 & 9.86 &  89.60 & 6.31  \\
UNET \cite{r27} & 28.9 & 88.10 & 9.91 & 89.30 & 6.30  \\ 
TransUnet \cite{r51} & 126.2 & 88.60 & 7.75 & 91.20 & 4.06  \\ 
PSPNet \cite{r52} & 48.8 & 88.00 & 10.11 & 89.30 & 6.29  \\ 
SegNet \cite{r53}  & 29.5 &  87.70 & 9.97 & 89.10 & 6.28 \\ 
ViT \cite{r54} & 85.6 & 88.20 & 8.92 & 90.80 & 5.58  \\ 
SwinTrans \cite{r55} & 88.3 &  88.80 & 7.41 &  91.10 & 4.68  \\ 
Segformer \cite{r56} & 64.1 & 88.40 & 7.56 &  90.60 & 5.89  \\ 
\midrule
\textbf{VesselSAM} &  \textbf{6.8} & \textbf{92.10} & \textbf{3.65} &  \textbf{93.20} & \textbf{3.10}  \\ 
\bottomrule
\end{tabular}
\end{minipage}
\hfill
\vspace*{0.02cm} 
    \newline
    \small
    \justifying
    Note: Bold indicates the best results.
\end{table*}

\subsection{Quantitative results}

To rigorously evaluate the performance of VesselSAM, we conducted a comprehensive quantitative comparison against a diverse set of state-of-the-art segmentation models. This evaluation includes models from three major categories: SAM-based models including SAM \cite{r6}, MedSAM \cite{r10}, SAMMed \cite{r9}, SAMed \cite{r11}, SAMAdp \cite{r19}; ViT-based models including UNETR \cite{r8}, TransUnet \cite{r51}, ViT \cite{r54}, SwinTrans \cite{r55}, SegFormer \cite{r56} and other Non-SAM-based methods including Xiong \cite{r35}, Sieren \cite{r36}, Li \cite{r37}, Zhao \cite{r38}, Feiger \cite{r39}, Wobben \cite{r40}, Song \cite{r41}, Hahn \cite{r42}, Abdolmanafi \cite{r43}, Chen \cite{r44}, Lyu \cite{r45}, Zhao \cite{r46}, Deng \cite{r47}, Yu \cite{r48}, Cheng \cite{r49}, Cao \cite{r50}, Unet \cite{r27}, PSPNet \cite{r52}, SegNet \cite{r53}. Each method was assessed under identical conditions to ensure a fair comparison, allowing us to accurately evaluate performance metrics such DSC and HD. The results demonstrate that VesselSAM surpasses existing SOTA models, effectively addressing challenges in complex medical image segmentation tasks.

\subsubsection{Quantitative Evaluation Results for AVT Dataset}

The performance metrics for various segmentation methods on the Aortic Vessel Tree (AVT) datasets, including Dongyang Hospital, Rider Hospital, and KiTs Hospital, are presented in \cref{tab1}. This comparison encompasses both big and small models, illustrating the effectiveness of each approach across multiple hospitals. VesselSAM demonstrates exceptional segmentation performance, achieving a DSC of 93.50\% at Dongyang Hospital, 93.25\% at Rider Hospital, and 93.02\% at Kits Hospital. This performance significantly surpasses that of state-of-the-art methods, including MedSAM and SAMAdp. 

The incorporation of Atrous Attention module and LoRA mechanisms within VesselSAM has contributed to its high performance, enabling the model to effectively capture multi-scale features essential for precise segmentation in medical imaging.  In contrast, models such as SAMMed and SAMed exhibit higher false positive rates, leading to suboptimal segmentation accuracy. This disparity underscores the advantages of VesselSAM in accurately delineating vascular structures amidst challenging imaging contexts, ultimately supporting its utility for clinical applications.

\subsubsection{Quantitative Evaluation Results for TBAD Dataset}

The results for the Type-B Aortic Dissection (TBAD) dataset are summarized in \cref{tab2}, further highlighting the performance of VesselSAM. The model achieves a DSC of 93.26\%, outperforming various competing methods, including UNETR and MedSAM. These findings illustrate VesselSAM's robustness in accurately segmenting the true lumen (TL) and false lumen (FL), emphasizing its effectiveness in handling complex segmentation tasks within clinical settings.
\par In comparison, SAM and MedSAM display lower performance, with DSC scores of 79.53\% and 92.20\%, respectively. Moreover, other models such as SAMMed and SAMAdp also exhibit challenges in segmentation accuracy, as evidenced by their lower DSC values. The consistent high performance of VesselSAM across both the AVT and TBAD datasets demonstrates its potential as a valuable tool for medical image segmentation, particularly in complex cases where precision is paramount.

\subsubsection{Applicability and Superiority over Non-SAM Benchmarks}

To rigorously assess the effectiveness of our SAM-based segmentation framework, VesselSAM, we conduct a comparative evaluation against a wide range of Non-SAM-based state-of-the-art segmentation benchmarks. As presented in \cref{tab3}, VesselSAM is compared with twenty-three~[34--57] existing Non-SAM methods specifically developed for aortic segmentation. Our method achieves a DSC of 92.10 and HD of 3.65 on the AVT dataset and a DSC of 93.20 with HD of 3.10 on TBAD dataset.

The consistently high DSC and low HD values across both datasets demonstrate the robustness, accuracy and generalization of our SAM-based approach. VesselSAM outperforms existing Non-SAM methods, establishing its superiority in handling diverse and independent datasets. These results validate the effectiveness of adapting SAM to the medical imaging domain, showcasing its potential as a reliable solution for vessel segmentation in real-world clinical scenarios.

\begin{figure*}[ht] 
    \includegraphics[width=\linewidth]{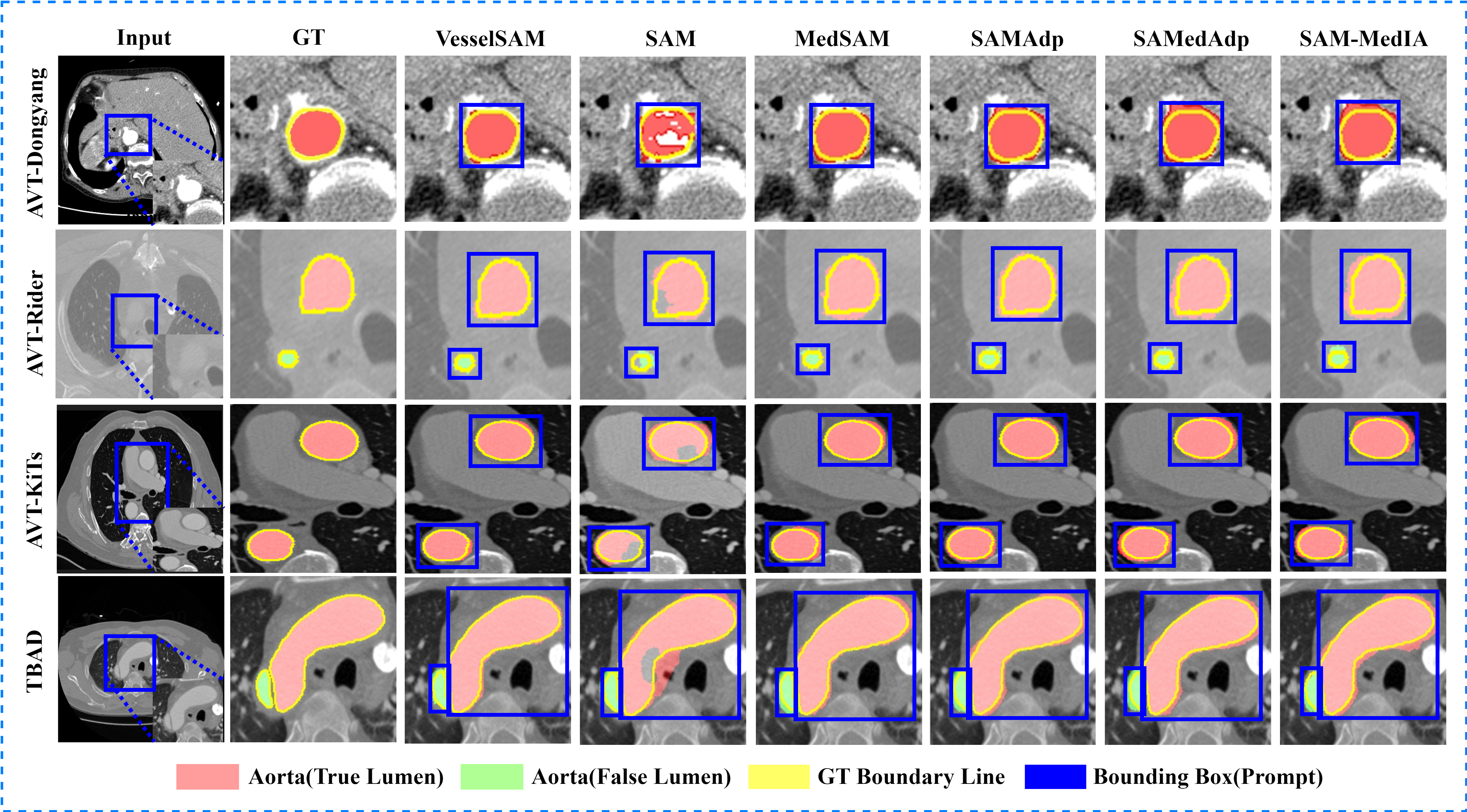}
    \caption{Qualitative visual results on the AVT (Dongyang, Rider, Kits) and TBAD Datasets under Bounding Box Prompts. The first column shows the input images, followed by the ground truth (GT) in the second column. The subsequent columns show the results of various segmentation models: VesselSAM, SAM, MedSAM, SAMAdp, SAMed, and SAMMed. Each model's output is overlaid with color-coded regions for true lumen (pink), false lumen (green), and the GT boundary line (yellow). The blue box represents the bounding box-prompt used for segmentation. The images have been zoomed in to enhance visibility.}
    \label{fig3}
\end{figure*}

\subsection{Qualitative results} 
To provide a more intuitive comparison, qualitative segmentation results are presented for VesselSAM and the same models evaluated in the quantitative analysis, as illustrated in \cref{fig3}. The top row displays the results for aortic vessel segmentation, while the bottom row highlights the segmentation of true lumen (TL) and false lumen (FL) for Type-B Aortic Dissection (TBAD). In the aortic vessel segmentation task, VesselSAM effectively delineates the vessel structures, capturing intricate details that may be overlooked by other models. The segmentation accurately follows the boundaries of the aorta, demonstrating its robustness in identifying the vessel amidst surrounding tissues. In contrast, SAM struggles with segmentation accuracy, leading to significant misalignments with the ground truth, particularly in the definition of vessel edges. MedSAM demonstrates improved performance compared to SAM but still fails to capture some fine details, resulting in inaccuracies in the vessel's contour. The models SAMMed, SAMed, and SAMAdp, struggles to accurately capture the true positive vessel areas, resulting in a significant number of false positive regions in their segmentations. While these models provide reasonable outputs, they tend to misidentify surrounding areas as part of the vessel structure.

In the segmentation of TL, FL and FLT in TBAD dataset, VesselSAM continues to excel by accurately capturing the luminal structures. The segmentation closely aligns with the GT, effectively distinguishing the TL, FL and the FLT. For better visualization, only the TL and FL are presented. In contrast, SAM encounters significant challenges, with poor segmentation of the TL, resulting in structural misrepresentations. MedSAM provides an improvement over SAM, but it still exhibits inaccuracies that affect its reliability in clinical applications. Other methods like SAMAdp, SAMed, and SAMMed similarly face challenges in accurately delineating the lumens, with occasional missing segments and imprecise boundaries.

\begin{figure*}[hb] 
    \centering
    \includegraphics[width=0.9\linewidth]{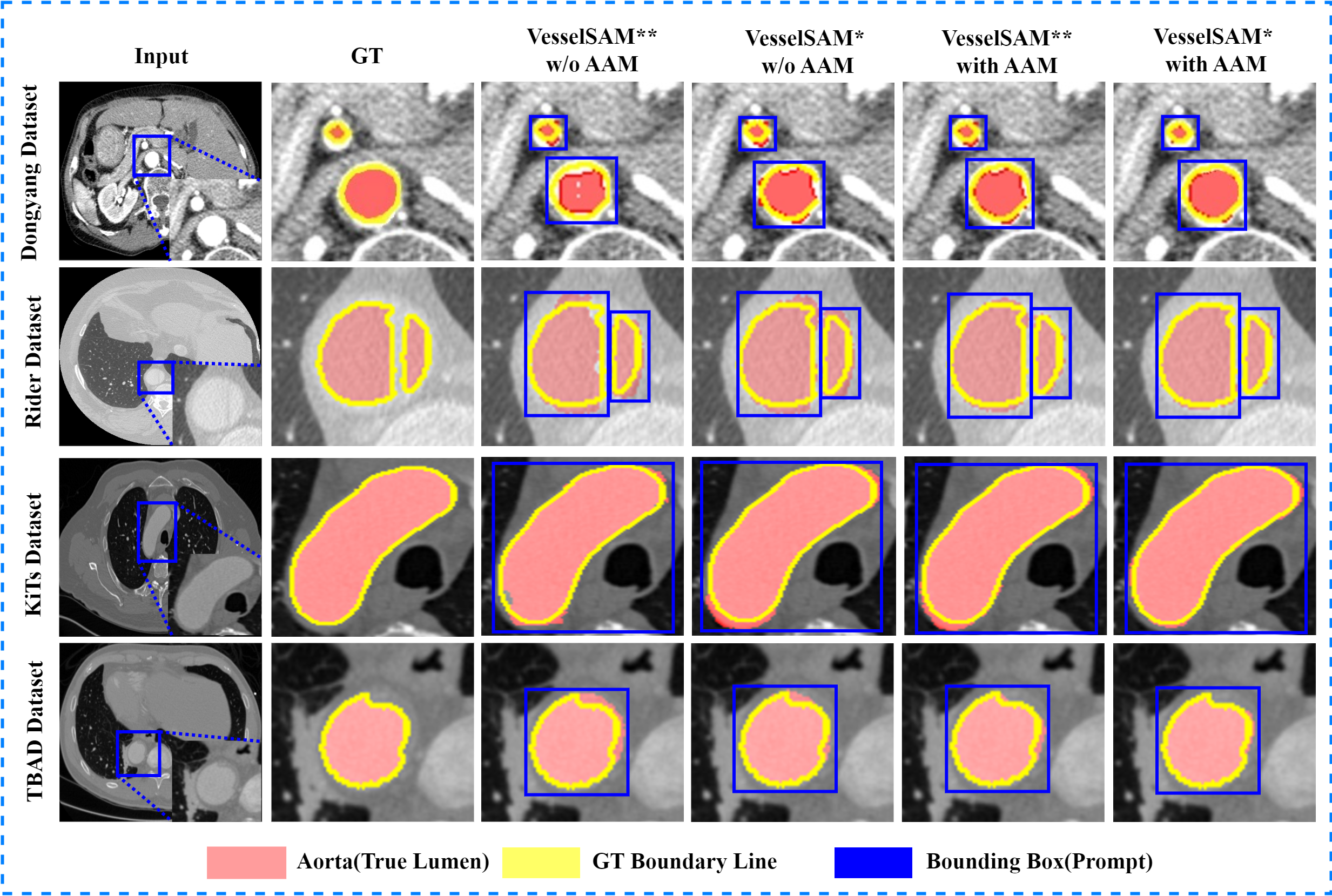}
    \caption{Qualitative visual results on the AVT (Dongyang, Rider, Kits) and TBAD datasets using bounding box prompts. The first column shows the input images, followed by the ground truth (GT) in the second column. Subsequent columns show the results of the following configurations: "VesselSAM* w/o AMM" refers to the VesselSAM model with the MedSAM backbone but without the Atrous Attention Module (AMM), while "VesselSAM** w/o AMM" employs the SAM backbone without AMM. "VesselSAM* with AMM" and "VesselSAM** with AMM" incorporate AMM with the MedSAM and SAM backbones, respectively. The outputs are color-coded to highlight the true lumen (pink), the GT boundary line (yellow), and the bounding box prompt (blue). The images have been zoomed in to enhance visibility.}
    \label{fig4}
\end{figure*}

\subsection{Ablation Study}

To evaluate the effectiveness of different configurations of the VesselSAM in medical image segmentation tasks, particularly vessel segmentation, we conducted a series of comprehensive ablation experiments. First, we compared the performance of two baseline models—VesselSAM initialized with the MedSAM (medical domain-specific) and SAM (general domain) configurations. Next, we tested an enhanced model incorporating the Atrous Attention module. The objective was to analyze the impact of these variations on segmentation performance, using the DSC as the primary evaluation metric.

\subsubsection{Impact of the Backbone and Atrous Attention Module}

To assess the impact of the backbone architecture and the integration of the Atrous Attention module on the performance of VesselSAM. We compare two configurations: VesselSAM initialized with the MedSAM backbone (VesselSAM*) and the SAM backbone (VesselSAM**), which serve as the baseline models for this analysis. Additionally, we introduce the Atrous Attention module into both configurations to evaluate its effect on segmentation performance.

\begin{table*}[ht]
    \centering
    \scriptsize
    \caption{Ablation study of Atrous Attention Module} \label{tab4}
    \begin{tabular}{p{2.8cm}p{2cm} p{2cm} p{2cm}p{2cm}}
        \toprule
        Dataset & VesselSAM* & VesselSAM** & AAM  & DSC \\
        \midrule
         & \ding{55}  & \checkmark  & \ding{55}   & 88.43\% \\
         & \checkmark  & \ding{55}  & \ding{55}   & 89.56\%  \\
        AVT-Dongyang \cite{r25} & \ding{55} & \checkmark  & \checkmark  & \underline{91.23\%}  \\
         & \checkmark  & \ding{55} & \checkmark  & \textbf{93.50\%} \\
        \midrule
         & \ding{55}  & \checkmark  & \ding{55}   & 88.25\% \\
         & \checkmark  & \ding{55}  & \ding{55}   & 89.16\%  \\
        AVT-KiTs \cite{r25}& \ding{55} & \checkmark  & \checkmark  & \underline{91.57\%}  \\
         & \checkmark  & \ding{55} & \checkmark  & \textbf{93.02\%} \\
        \midrule
         & \ding{55}  & \checkmark  & \ding{55}   & 87.89\% \\
         & \checkmark  & \ding{55}  & \ding{55}   & 88.75\%  \\
        AVT-Rider \cite{r25} & \ding{55} & \checkmark  & \checkmark  & \underline{91.42\%}  \\
         & \checkmark  & \ding{55} & \checkmark  & \textbf{93.25\%} \\
        \midrule
         & \ding{55}  & \checkmark  & \ding{55}   & 90.76\% \\
         & \checkmark  & \ding{55}  & \ding{55}   & 91.68\%  \\
        TBAD \cite{r26} & \ding{55} & \checkmark  & \checkmark  & \underline{91.90\%}  \\
         & \checkmark  & \ding{55} & \checkmark  & \textbf{93.26\%} \\
        \bottomrule
    \end{tabular}%
\vspace*{0.05cm} 
\par 
\raggedright
\small Note:  Bold indicates the best results and underline denotes the second best results. VesselSAM* represents the VesselSAM Model with the MedSAM Model as a backbone, VesselSAM** represents the VesselSAM Model with the SAM Model as a backbone and AAM represents Atrous Attention Module. 
\end{table*}

\begin{figure}[ht]
\centering
\begin{tikzpicture}
\begin{axis}[
    xlabel={Epochs},
    ylabel={Training Loss},
    xmin=0, xmax=52,
    ymin=0, ymax=0.013,
    xtick={0,10,20,30,40,50},
    ytick={0.002,0.004,0.006,0.008,0.010,0.012},
    yticklabels={0.002,0.004,0.006,0.008,0.010,0.012},
    yticklabel style={/pgf/number format/fixed, /pgf/number format/precision=4},
    scaled y ticks=false, 
    legend style={at={(0.98,0.98)}, anchor=north east, draw=gray, fill=white, thick, draw opacity=0.7, text opacity=1},
    axis lines=box, 
    xtick pos=bottom, 
    ytick pos=left,
    ]

\addplot[
    color=red,
    mark=x,
    ]
    coordinates {
    (0,0.010)(1,0.009)(2,0.008)(5,0.006)(10,0.004)(15,0.0035)(20,0.0028)(25,0.0025)(30,0.0023)(35,0.0021)(40,0.002)(45,0.002)(50,0.002)
    };
    \addlegendentry{w Atrous Attention }

\addplot[
    color=green,
    mark=o,
    ]
    coordinates {
    (0,0.011)(1,0.0095)(2,0.009)(5,0.0075)(10,0.006)(15,0.004)(20,0.0035)(25,0.003)(30,0.0028)(35,0.003)(40,0.0025)(45,0.0023)(50,0.0024)
    };
    \addlegendentry{w/o Atrous Attention }

    \draw[-{Stealth[scale=1.5]}] (axis cs:22,0.0055) -- (axis cs:20,0.0025)
node[pos=0, align=center, above] {Converge faster and\\better!};
\end{axis}
\end{tikzpicture}
\caption{The impact of Atrous Attention Module on validation loss over training epochs.}
\label{fig5}
\end{figure}

The Atrous Attention module is integrated into the image encoder to improve the model's ability to capture multi-scale features. By utilizing dilated convolutions, this module expands the receptive field, enabling the model to focus on both small and large structures within the input image. This is particularly important for accurately segmenting vascular structures, where both fine details and broader contextual information are essential.

From the results presented in \cref{fig4}, it is evident that the Atrous Attention module improves the segmentation accuracy of both the MedSAM and SAM backbones. The segmentation outputs, which highlight the true lumen (pink), the GT boundary line (yellow), and the bounding box prompt (blue), demonstrate enhanced delineation of vascular structures when the Atrous Attention module is applied.

The quantitative results in \cref{tab4} provide strong evidence supporting the effectiveness of integrating the Atrous Attention module with the MedSAM backbone. The configuration combining the MedSAM backbone with the Atrous Attention module (VesselSAM* with AAM) achieved the highest Dice score of 93.50\% on the AVT-Dongyang dataset, outperforming all other configurations. This result highlights the significant benefit of using the MedSAM backbone, specifically designed for medical imaging, in combination with the Atrous Attention module, which enhances the model’s ability to capture multi-scale features. This combination provides a substantial improvement in segmentation accuracy, making it the most effective configuration for vascular segmentation.

In comparison, the VesselSAM model with the SAM backbone (VesselSAM** with AAM) also benefits from the Atrous Attention module, but the Dice scores are consistently lower. While these results still reflect an improvement over the baseline model with the Atrous Attention module, they demonstrate that the MedSAM backbone tailored for medical applications, offers a clear advantage when combined with the Atrous Attention module.

These findings suggest that the Atrous Attention module consistently improves segmentation performance, but its full potential is realized when paired with a domain-specific backbone like MedSAM. This combination enables VesselSAM to achieve the best performance across multiple datasets, reinforcing the importance of both the backbone architecture and attention mechanism in improving segmentation accuracy.

The training dynamics are further illustrated in \cref{fig5}, where the training loss curves for both configurations are compared. The model with Atrous Attention (red line) shows faster convergence and lower validation loss compared to the model without the Atrous Attention module (green line). By around epoch 20, the model with Atrous Attention stabilizes at a lower training loss, indicating that the module accelerates convergence and enhances the model’s ability to segment vascular structures more accurately.

\subsubsection{Impact of LoRA Rank}

In this experiment, we investigated the effect of LoRA rank on the performance of VesselSAM. Low-Rank Adaptation (LoRA) is designed to reduce the number of trainable parameters, making the training process more efficient without compromising model performance. We tested different LoRA ranks (2, 4, 16, 32, and 64) and measured their impact on segmentation accuracy using the Dice score as the evaluation metric.

\begin{figure}[ht]
	\centering
	\includegraphics[width=0.6\linewidth]{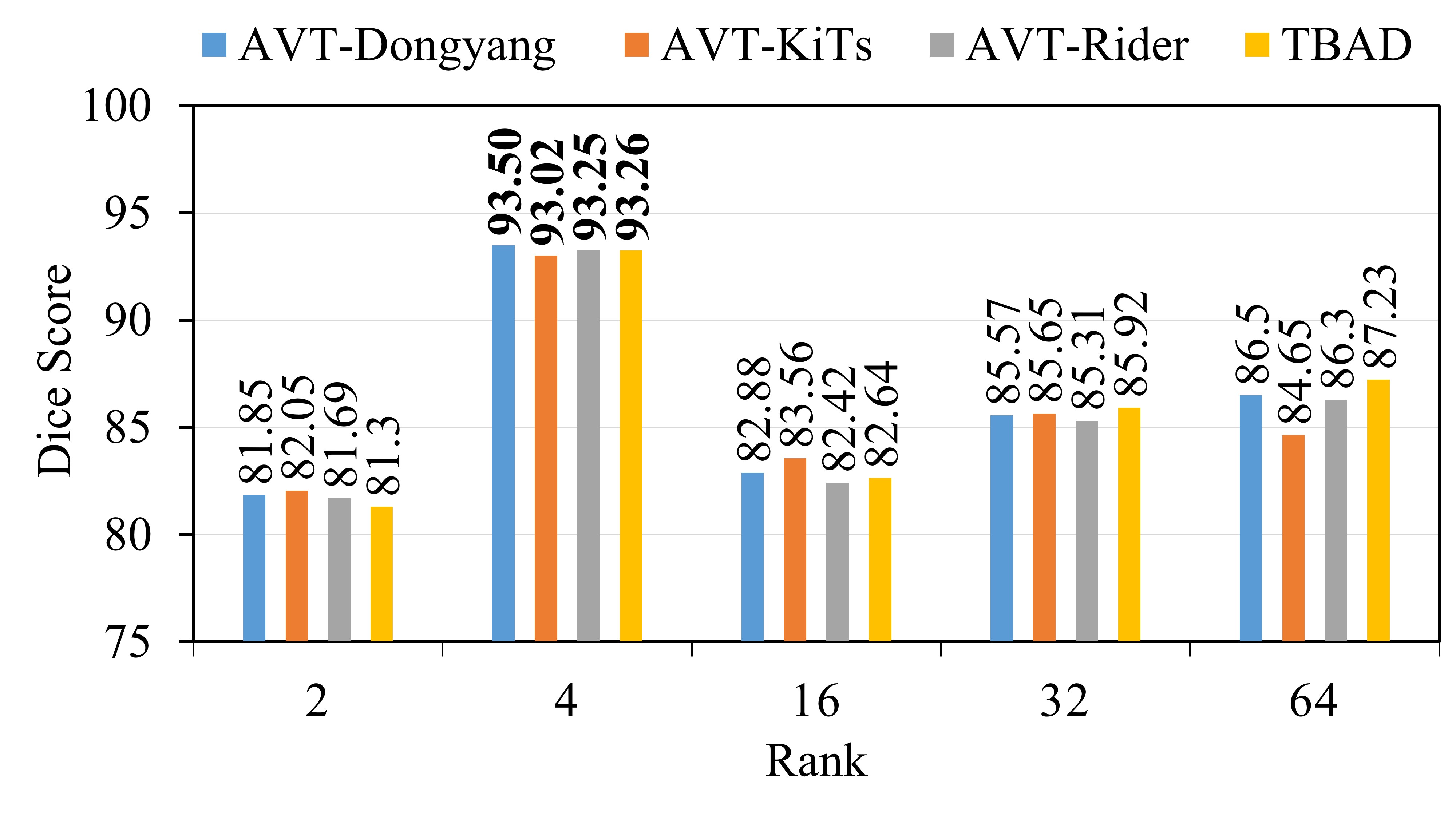}
	\caption{DSC VS Rank: Comparison of Dice Similarity Coefficients (DSC) for Aortic Vascular Tree (AVT) datasets (Dongyang, KiTs, Rider) and TBAD dataset across different LoRA ranks (2, 4, 16, 32, and 64), illustrating the performance stability and optimal rank selection for segmentation tasks.}
	\label{fig6}
\end{figure}

As illustrated in \cref{fig6}, the performance of VesselSAM showed significant variation across different LoRA ranks. LoRA rank 4 yielded the best performance, with the model achieving a Dice score of 93.5 on the AVT-Dongyang dataset, and similar strong performance on other datasets: 93.02 on AVT-KiTs, 93.25 on AVT-Rider, and 93.26 on TBAD. This suggests that LoRA rank 4 offers the optimal trade-off between segmentation accuracy and computational efficiency.
However, as the LoRA rank increased beyond 4, performance started to decline. For instance, at LoRA rank 16, the AVT-Dongyang Dice score dropped to 82.88, and at LoRA rank 32, it further decreased to 85.57. Interestingly, LoRA rank 64 resulted in slightly improved scores compared to rank 32, but still did not outperform rank 4. This trend indicates diminishing returns as the LoRA rank increases beyond an optimal point, with rank 4 providing the best overall segmentation performance.

\subsection{Limitations and Future Work}
This study demonstrates that domain-specific models, such as MedSAM, can achieve superior segmentation accuracy when enhanced with adaptation techniques like AtrousLoRA and Atrous Attention Module. These findings highlight the importance of parameter efficient fine-tuning strategies for optimizing medical image segmentation, particularly under computational constraints. Despite its strong performance in aortic vessel segmentation, VesselSAM has several limitations that warrant further investigation. One primary limitation is its reliance on bounding box prompts, which may not always provide sufficient contextual information for segmenting complex or ambiguous vascular structures. To improve flexibility and accuracy, future work will explore alternative prompt mechanisms, such as text-based prompts, to offer richer, more intuitive guidance for segmentation tasks.

Another challenge is VesselSAM’s dependency on high-quality input images. While the model performs well on clean, well-annotated datasets, its segmentation accuracy may degrade in noisy, low-resolution, or real-world clinical imaging conditions. To address this, future research will focus on enhancing the model’s robustness through advanced data augmentation techniques and strategies to improve generalization across diverse medical imaging domains. 

Furthermore, the integration of visual-language models (VLMs) with VesselSAM presents an exciting direction for future work. By leveraging language-driven prompts, these models could refine segmentation accuracy and enable the system to handle ambiguous or novel vascular structures with minimal user input. Additionally, expanding VesselSAM’s applicability beyond aortic vessel segmentation is crucial. Investigating its performance on other vascular structures, such as coronary arteries, cerebral vessels, and peripheral vasculature, could further enhance its clinical utility across multiple medical domains. By addressing these limitations and exploring these future directions, VesselSAM can evolve into a more generalized, adaptive, and clinically impactful segmentation framework for medical image analysis.

\section{Conclusion}
In this paper, we introduced VesselSAM, an enhanced adaptation of the Segment Anything Model (SAM), specifically designed for aortic vessel segmentation. By integrating AtrousLoRA, a novel combination of Atrous Attention and Low-Rank Adaptation (LoRA), VesselSAM effectively overcomes key limitations of the original SAM, improving its ability to capture complex hierarchical features in medical images. The Atrous Attention Module facilitates multi-scale feature extraction, preserving both fine-grained details and broader anatomical structures, while LoRA optimizes fine-tuning efficiency, significantly reducing trainable parameters without compromising segmentation accuracy.

Extensive evaluations on the Aortic Vessel Tree (AVT) and Type-B Aortic Dissection (TBAD) datasets demonstrate that VesselSAM outperforms state-of-the-art (SOTA) ViT-based and SAM-based models, achieving superior DSC and Hausdorff Distance HD scores. Notably, VesselSAM achieves these results with fewer trainable parameters, reinforcing its position as an efficient Parameter-Efficient Fine-Tuning (PEFT) model for medical imaging applications. These findings highlight its ability to deliver high segmentation accuracy while maintaining computational efficiency, making it highly valuable for real-world clinical deployment.

The VesselSAM offers a robust and scalable solution for vascular image segmentation, demonstrating strong generalization across diverse vascular datasets while maintaining minimal computational overhead. Future work will focus on further enhancing its adaptability, including the integration of text-based prompts and visual language models to enrich segmentation guidance, as well as extending its applicability to other vascular structures and medical imaging tasks. These advancements will further solidify VesselSAM’s role as a versatile and efficient AI-driven tool for clinical and research applications in medical imaging.

\section{Declaration of Competing Interest}
The authors declare that they have no conflicts of interest.

\section{Acknowledgements}
This work was supported by the Natural Science Foundation of Guangdong Province (No. 2023A1515010673), in part by the Shenzhen Science and Technology Innovation Bureau key project (No. JSGG20220831110400001, No. CJGJZD20230724093303007, KJZD20240903101259001), in part by Shenzhen Medical Research Fund (No. D2404001), in part by Shenzhen Engineering Laboratory for Diagnosis \& Treatment Key Technologies of Interventional Surgical Robots (XMHT20220104009), and the Key Laboratory of Biomedical Imaging Science and System, CAS  and
the University Chinese Academy of Sciences and Alliance of International Science Organization (ANSO) through 2021A8017729012.

\bibliographystyle{model1-num-names}
\bibliography{main}

\begin{thebibliography}{56}
\expandafter\ifx\csname natexlab\endcsname\relax\def\natexlab#1{#1}\fi
\providecommand{\url}[1]{\texttt{#1}}
\providecommand{\href}[2]{#2}
\providecommand{\path}[1]{#1}
\providecommand{\DOIprefix}{doi:}
\providecommand{\ArXivprefix}{arXiv:}
\providecommand{\URLprefix}{URL: }
\providecommand{\Pubmedprefix}{pmid:}
\providecommand{\doi}[1]{\href{http://dx.doi.org/#1}{\path{#1}}}
\providecommand{\Pubmed}[1]{\href{pmid:#1}{\path{#1}}}
\providecommand{\bibinfo}[2]{#2}
\ifx\xfnm\relax \def\xfnm[#1]{\unskip,\space#1}\fi
\bibitem[{Li et~al.(2023)Li, Jiang, Zhang, and Zhu}]{r1}
\bibinfo{author}{M.~Li}, \bibinfo{author}{Y.~Jiang}, \bibinfo{author}{Y.~Zhang}, \bibinfo{author}{H.~Zhu},
\newblock \bibinfo{title}{Medical image analysis using deep learning algorithms},
\newblock \bibinfo{journal}{Front. Public Health} \bibinfo{volume}{11} (\bibinfo{year}{2023}) \bibinfo{pages}{1273253}.
\bibitem[{Jin et~al.(2021)Jin, Pepe, Li, Gsaxner, Zhao, Pomykala, Kleesiek, Frangi, and Egger}]{r2}
\bibinfo{author}{Y.~Jin}, \bibinfo{author}{A.~Pepe}, \bibinfo{author}{J.~Li}, \bibinfo{author}{C.~Gsaxner}, \bibinfo{author}{F.~Zhao}, \bibinfo{author}{K.~L. Pomykala}, \bibinfo{author}{J.~Kleesiek}, \bibinfo{author}{A.~F. Frangi}, \bibinfo{author}{J.~Egger},
\newblock \bibinfo{title}{Ai-based aortic vessel tree segmentation for cardiovascular diseases treatment: status quo},
\newblock \bibinfo{journal}{arXiv preprint arXiv:2108.02998}  (\bibinfo{year}{2021}).
\bibitem[{Xiao et~al.(2023)Xiao, Li, Liu, Zhu, and Zhang}]{r3}
\bibinfo{author}{H.~Xiao}, \bibinfo{author}{L.~Li}, \bibinfo{author}{Q.~Liu}, \bibinfo{author}{X.~Zhu}, \bibinfo{author}{Q.~Zhang},
\newblock \bibinfo{title}{Transformers in medical image segmentation: A review},
\newblock \bibinfo{journal}{Biomedical Signal Processing and Control} \bibinfo{volume}{84} (\bibinfo{year}{2023}) \bibinfo{pages}{104791}.
\bibitem[{Han et~al.(2022)Han, Wang, Chen, Chen, Guo, Liu, Tang, Xiao, Xu, Xu, and Yang}]{r4}
\bibinfo{author}{K.~Han}, \bibinfo{author}{Y.~Wang}, \bibinfo{author}{H.~Chen}, \bibinfo{author}{X.~Chen}, \bibinfo{author}{J.~Guo}, \bibinfo{author}{Z.~Liu}, \bibinfo{author}{Y.~Tang}, \bibinfo{author}{A.~Xiao}, \bibinfo{author}{C.~Xu}, \bibinfo{author}{Y.~Xu}, \bibinfo{author}{Z.~Yang},
\newblock \bibinfo{title}{A survey on vision transformer},
\newblock \bibinfo{journal}{IEEE Trans. Pattern Anal. Mach. Intell.} \bibinfo{volume}{45} (\bibinfo{year}{2022}) \bibinfo{pages}{87--110}.
\bibitem[{Li et~al.(2024)Li, Li, Sun, and Weng}]{r5}
\bibinfo{author}{S.~Li}, \bibinfo{author}{B.~Li}, \bibinfo{author}{B.~Sun}, \bibinfo{author}{Y.~Weng},
\newblock \bibinfo{title}{Towards visual-prompt temporal answer grounding in instructional video},
\newblock \bibinfo{journal}{IEEE Trans. Pattern Anal. Mach. Intell.} \bibinfo{volume}{46} (\bibinfo{year}{2024}) \bibinfo{pages}{8836--8853}.
\bibitem[{Kirillov et~al.(2023)Kirillov, Mintun, Ravi, Mao, Rolland, Gustafson, Xiao, Whitehead, Berg, Lo et~al.}]{r6}
\bibinfo{author}{A.~Kirillov}, \bibinfo{author}{E.~Mintun}, \bibinfo{author}{N.~Ravi}, \bibinfo{author}{H.~Mao}, \bibinfo{author}{C.~Rolland}, \bibinfo{author}{L.~Gustafson}, \bibinfo{author}{T.~Xiao}, \bibinfo{author}{S.~Whitehead}, \bibinfo{author}{A.~C. Berg}, \bibinfo{author}{W.-Y. Lo}, et~al.,
\newblock \bibinfo{title}{Segment anything},
\newblock in: \bibinfo{booktitle}{Proceedings of the IEEE/CVF international conference on computer vision}, \bibinfo{year}{2023}, pp. \bibinfo{pages}{4015--4026}.
\bibitem[{Cao et~al.(2022)Cao, Wang, Chen, Jiang, Zhang, Tian, and Wang}]{r7}
\bibinfo{author}{H.~Cao}, \bibinfo{author}{Y.~Wang}, \bibinfo{author}{J.~Chen}, \bibinfo{author}{D.~Jiang}, \bibinfo{author}{X.~Zhang}, \bibinfo{author}{Q.~Tian}, \bibinfo{author}{M.~Wang},
\newblock \bibinfo{title}{Swin-unet: Unet-like pure transformer for medical image segmentation},
\newblock \bibinfo{journal}{European conference on computer vision}  (\bibinfo{year}{2022}) \bibinfo{pages}{205--218}.
\bibitem[{Hatamizadeh et~al.(2022)Hatamizadeh, Tang, Nath, Yang, Myronenko, Landman, Roth, and Xu}]{r8}
\bibinfo{author}{A.~Hatamizadeh}, \bibinfo{author}{Y.~Tang}, \bibinfo{author}{V.~Nath}, \bibinfo{author}{D.~Yang}, \bibinfo{author}{A.~Myronenko}, \bibinfo{author}{B.~Landman}, \bibinfo{author}{H.~R. Roth}, \bibinfo{author}{D.~Xu},
\newblock \bibinfo{title}{Unetr: Transformers for 3d medical image segmentation},
\newblock in: \bibinfo{booktitle}{Proceedings of the IEEE/CVF winter conference on applications of computer vision}, \bibinfo{year}{2022}, pp. \bibinfo{pages}{574--584}.
\bibitem[{Huang et~al.(2024)Huang, Yang, Liu, Zhou, Chang, Zhou, Chen, Yu, Chen, Chen et~al.}]{r9}
\bibinfo{author}{Y.~Huang}, \bibinfo{author}{X.~Yang}, \bibinfo{author}{L.~Liu}, \bibinfo{author}{H.~Zhou}, \bibinfo{author}{A.~Chang}, \bibinfo{author}{X.~Zhou}, \bibinfo{author}{R.~Chen}, \bibinfo{author}{J.~Yu}, \bibinfo{author}{J.~Chen}, \bibinfo{author}{C.~Chen}, et~al.,
\newblock \bibinfo{title}{Segment anything model for medical images?},
\newblock \bibinfo{journal}{Medical Image Analysis} \bibinfo{volume}{92} (\bibinfo{year}{2024}) \bibinfo{pages}{103061}.
\bibitem[{Ma et~al.(2024)Ma, He, Li, Han, You, and Wang}]{r10}
\bibinfo{author}{J.~Ma}, \bibinfo{author}{Y.~He}, \bibinfo{author}{F.~Li}, \bibinfo{author}{L.~Han}, \bibinfo{author}{C.~You}, \bibinfo{author}{B.~Wang},
\newblock \bibinfo{title}{Segment anything in medical images},
\newblock \bibinfo{journal}{Nature Communications} \bibinfo{volume}{15} (\bibinfo{year}{2024}) \bibinfo{pages}{654}.
\bibitem[{Zhang and Liu(2023)}]{r11}
\bibinfo{author}{K.~Zhang}, \bibinfo{author}{D.~Liu},
\newblock \bibinfo{title}{Customized segment anything model for medical image segmentation},
\newblock \bibinfo{journal}{arXiv preprint arXiv:2304.13785}  (\bibinfo{year}{2023}).
\bibitem[{Mazurowski et~al.(2023)Mazurowski, Dong, Gu, Yang, Konz, and Zhang}]{r12}
\bibinfo{author}{M.~A. Mazurowski}, \bibinfo{author}{H.~Dong}, \bibinfo{author}{H.~Gu}, \bibinfo{author}{J.~Yang}, \bibinfo{author}{N.~Konz}, \bibinfo{author}{Y.~Zhang},
\newblock \bibinfo{title}{Segment anything model for medical image analysis: an experimental study},
\newblock \bibinfo{journal}{Medical Image Analysis} \bibinfo{volume}{89} (\bibinfo{year}{2023}) \bibinfo{pages}{102918}.
\bibitem[{Zhang et~al.(2024)Zhang, Shen, and Jiao}]{r13}
\bibinfo{author}{Y.~Zhang}, \bibinfo{author}{Z.~Shen}, \bibinfo{author}{R.~Jiao},
\newblock \bibinfo{title}{Segment anything model for medical image segmentation: Current applications and future directions},
\newblock \bibinfo{journal}{Computers in Biology and Medicine}  (\bibinfo{year}{2024}) \bibinfo{pages}{108238}.
\bibitem[{Deng et~al.(2023)Deng, Cui, Liu, Yao, Remedios, Bao, Landman, Wheless, Coburn, Wilson et~al.}]{r14}
\bibinfo{author}{R.~Deng}, \bibinfo{author}{C.~Cui}, \bibinfo{author}{Q.~Liu}, \bibinfo{author}{T.~Yao}, \bibinfo{author}{L.~W. Remedios}, \bibinfo{author}{S.~Bao}, \bibinfo{author}{B.~A. Landman}, \bibinfo{author}{L.~E. Wheless}, \bibinfo{author}{L.~A. Coburn}, \bibinfo{author}{K.~T. Wilson}, et~al.,
\newblock \bibinfo{title}{Segment anything model (sam) for digital pathology: Assess zero-shot segmentation on whole slide imaging},
\newblock \bibinfo{journal}{arXiv preprint arXiv:2304.04155}  (\bibinfo{year}{2023}).
\bibitem[{Hu et~al.(2023)Hu, Xia, Ju, and Li}]{r15}
\bibinfo{author}{C.~Hu}, \bibinfo{author}{T.~Xia}, \bibinfo{author}{S.~Ju}, \bibinfo{author}{X.~Li},
\newblock \bibinfo{title}{When sam meets medical images: An investigation of segment anything model (sam) on multi-phase liver tumor segmentation},
\newblock \bibinfo{journal}{arXiv preprint arXiv:2304.08506}  (\bibinfo{year}{2023}).
\bibitem[{He et~al.(2023)He, Bao, Li, Stout, Bjornerud, Grant, and Ou}]{r16}
\bibinfo{author}{S.~He}, \bibinfo{author}{R.~Bao}, \bibinfo{author}{J.~Li}, \bibinfo{author}{J.~Stout}, \bibinfo{author}{A.~Bjornerud}, \bibinfo{author}{P.~E. Grant}, \bibinfo{author}{Y.~Ou},
\newblock \bibinfo{title}{Computer-vision benchmark segment-anything model (sam) in medical images: Accuracy in 12 datasets},
\newblock \bibinfo{journal}{arXiv preprint arXiv:2304.09324}  (\bibinfo{year}{2023}).
\bibitem[{Hu et~al.(2021)Hu, Shen, Wallis, Allen-Zhu, Li, Wang, Wang, and Chen}]{r17}
\bibinfo{author}{E.~J. Hu}, \bibinfo{author}{Y.~Shen}, \bibinfo{author}{P.~Wallis}, \bibinfo{author}{Z.~Allen-Zhu}, \bibinfo{author}{Y.~Li}, \bibinfo{author}{S.~Wang}, \bibinfo{author}{L.~Wang}, \bibinfo{author}{W.~Chen},
\newblock \bibinfo{title}{Lora: Low-rank adaptation of large language models},
\newblock \bibinfo{journal}{arXiv preprint arXiv:2106.09685}  (\bibinfo{year}{2021}).
\bibitem[{Li and Rajpurkar(2024)}]{r18}
\bibinfo{author}{K.~Li}, \bibinfo{author}{P.~Rajpurkar},
\newblock \bibinfo{title}{Adapting segment anything models to medical imaging via fine-tuning without domain pretraining},
\newblock \bibinfo{journal}{AAAI 2024 Spring Symposium on Clinical Foundation Models}  (\bibinfo{year}{2024}).
\bibitem[{Chen et~al.(2023)Chen, Zhu, Deng, Cao, Wang, Zhang, Li, Sun, Zang, and Mao}]{r19}
\bibinfo{author}{T.~Chen}, \bibinfo{author}{L.~Zhu}, \bibinfo{author}{C.~Deng}, \bibinfo{author}{R.~Cao}, \bibinfo{author}{Y.~Wang}, \bibinfo{author}{S.~Zhang}, \bibinfo{author}{Z.~Li}, \bibinfo{author}{L.~Sun}, \bibinfo{author}{Y.~Zang}, \bibinfo{author}{P.~Mao},
\newblock \bibinfo{title}{Sam-adapter: Adapting segment anything in underperformed scenes},
\newblock in: \bibinfo{booktitle}{Proceedings of the IEEE/CVF International Conference on Computer Vision}, \bibinfo{year}{2023}, pp. \bibinfo{pages}{3367--3375}.
\bibitem[{Chen et~al.(2022)Chen, Duan, Wang, He, Lu, Dai, and Qiao}]{r20}
\bibinfo{author}{Z.~Chen}, \bibinfo{author}{Y.~Duan}, \bibinfo{author}{W.~Wang}, \bibinfo{author}{J.~He}, \bibinfo{author}{T.~Lu}, \bibinfo{author}{J.~Dai}, \bibinfo{author}{Y.~Qiao},
\newblock \bibinfo{title}{Vision transformer adapter for dense predictions},
\newblock \bibinfo{journal}{The Eleventh International Conference on Learning Representations.}  (\bibinfo{year}{2022}).
\bibitem[{Chen et~al.(2017)Chen, Papandreou, Kokkinos, Murphy, and Yuille}]{r21}
\bibinfo{author}{L.-C. Chen}, \bibinfo{author}{G.~Papandreou}, \bibinfo{author}{I.~Kokkinos}, \bibinfo{author}{K.~Murphy}, \bibinfo{author}{A.~L. Yuille},
\newblock \bibinfo{title}{Deeplab: Semantic image segmentation with deep convolutional nets, atrous convolution, and fully connected crfs},
\newblock \bibinfo{journal}{IEEE Trans. Pattern Anal. Mach. Intell.} \bibinfo{volume}{40} (\bibinfo{year}{2017}) \bibinfo{pages}{834--848}.
\bibitem[{Hu et~al.(2025)Hu, Li, Jain, Lin, and Chen}]{r22}
\bibinfo{author}{J.~Hu}, \bibinfo{author}{Y.~Li}, \bibinfo{author}{R.~K. Jain}, \bibinfo{author}{L.~Lin}, \bibinfo{author}{Y.~Chen},
\newblock \bibinfo{title}{Spa: Leveraging the sam with spatial priors adapter for enhanced medical image segmentation},
\newblock \bibinfo{journal}{IEEE J. Biomed. Health Inform.}  (\bibinfo{year}{2025}).
\bibitem[{Wu et~al.(2021)Wu, Xiao, Codella, Liu, Dai, Yuan, and Zhang}]{r23}
\bibinfo{author}{H.~Wu}, \bibinfo{author}{B.~Xiao}, \bibinfo{author}{N.~Codella}, \bibinfo{author}{M.~Liu}, \bibinfo{author}{X.~Dai}, \bibinfo{author}{L.~Yuan}, \bibinfo{author}{L.~Zhang},
\newblock \bibinfo{title}{Cvt: Introducing convolutions to vision transformers},
\newblock in: \bibinfo{booktitle}{Proceedings of the IEEE/CVF international conference on computer vision}, \bibinfo{year}{2021}, pp. \bibinfo{pages}{22--31}.
\bibitem[{Ibtehaz et~al.(2024)Ibtehaz, Yan, Mortazavi, and Kihara}]{r24}
\bibinfo{author}{N.~Ibtehaz}, \bibinfo{author}{N.~Yan}, \bibinfo{author}{M.~Mortazavi}, \bibinfo{author}{D.~Kihara},
\newblock \bibinfo{title}{Acc-vit: Atrous convolution's comeback in vision transformers},
\newblock \bibinfo{journal}{arXiv preprint arXiv:2403.04200}  (\bibinfo{year}{2024}).
\bibitem[{Liu et~al.(2022)Liu, Fan, and Zhou}]{Liu2022}
\bibinfo{author}{X.~Liu}, \bibinfo{author}{W.~Fan}, \bibinfo{author}{D.~Zhou},
\newblock \bibinfo{title}{Skin lesion segmentation via intensive atrous spatial transformer},
\newblock in: \bibinfo{booktitle}{International Conference on Wireless Algorithms, Systems, and Applications}, \bibinfo{publisher}{Springer Nature Switzerland}, \bibinfo{year}{2022}, pp. \bibinfo{pages}{15--26}.
\bibitem[{Tong et~al.(2024)Tong, Li, Zhang, Zhang, Zhu, Du, and Hu}]{TONG2024213}
\bibinfo{author}{L.~Tong}, \bibinfo{author}{T.~Li}, \bibinfo{author}{Q.~Zhang}, \bibinfo{author}{Q.~Zhang}, \bibinfo{author}{R.~Zhu}, \bibinfo{author}{W.~Du}, \bibinfo{author}{P.~Hu},
\newblock \bibinfo{title}{Livit-net: A u-net-like, lightweight transformer network for retinal vessel segmentation},
\newblock \bibinfo{journal}{Computational and Structural Biotechnology}  (\bibinfo{year}{2024}) \bibinfo{pages}{213--224}.
\bibitem[{Lam et~al.(2021)Lam, Lim, Sutopo, and Baskaran}]{Lam2021}
\bibinfo{author}{A.~Lam}, \bibinfo{author}{J.~Y. Lim}, \bibinfo{author}{R.~Sutopo}, \bibinfo{author}{V.~M. Baskaran},
\newblock \bibinfo{title}{Paying attention to varying receptive fields: object detection with atrous filters and vision transformers},
\newblock in: \bibinfo{booktitle}{British Machine Vision Conference 2021}, \bibinfo{organization}{British Machine Vision Association}, \bibinfo{year}{2021}.
\bibitem[{Yu and Koltun(2015)}]{Yu2015}
\bibinfo{author}{F.~Yu}, \bibinfo{author}{V.~Koltun},
\newblock \bibinfo{title}{Multi-scale context aggregation by dilated convolutions},
\newblock \bibinfo{journal}{arXiv preprint arXiv:1511.07122}  (\bibinfo{year}{2015}).
\bibitem[{Agarap(2018)}]{agarap2018deep}
\bibinfo{author}{A.~F. Agarap},
\newblock \bibinfo{title}{Deep learning using rectified linear units (relu)},
\newblock \bibinfo{journal}{arXiv preprint arXiv:1803.08375}  (\bibinfo{year}{2018}).
\bibitem[{Ioffe and Szegedy(2015)}]{ioffe2015batch}
\bibinfo{author}{S.~Ioffe}, \bibinfo{author}{C.~Szegedy},
\newblock \bibinfo{title}{Batch normalization: Accelerating deep network training by reducing internal covariate shift},
\newblock in: \bibinfo{booktitle}{International conference on machine learning}, \bibinfo{organization}{pmlr}, \bibinfo{year}{2015}, pp. \bibinfo{pages}{448--456}.
\bibitem[{Radl et~al.(2022)Radl, Jin, Pepe, Li, Gsaxner, Zhao, and Egger}]{r25}
\bibinfo{author}{L.~Radl}, \bibinfo{author}{Y.~Jin}, \bibinfo{author}{A.~Pepe}, \bibinfo{author}{J.~Li}, \bibinfo{author}{C.~Gsaxner}, \bibinfo{author}{F.~Zhao}, \bibinfo{author}{J.~Egger},
\newblock \bibinfo{title}{Avt: Multicenter aortic vessel tree cta dataset collection with ground truth segmentation masks},
\newblock \bibinfo{journal}{Data in brief} \bibinfo{volume}{40} (\bibinfo{year}{2022}) \bibinfo{pages}{107801}.
\bibitem[{Yao et~al.(2021)Yao, Xie, Zhang, Dong, Qiu, Yuan, Jia, Wang, Shi, Zhuang et~al.}]{r26}
\bibinfo{author}{Z.~Yao}, \bibinfo{author}{W.~Xie}, \bibinfo{author}{J.~Zhang}, \bibinfo{author}{Y.~Dong}, \bibinfo{author}{H.~Qiu}, \bibinfo{author}{H.~Yuan}, \bibinfo{author}{Q.~Jia}, \bibinfo{author}{T.~Wang}, \bibinfo{author}{Y.~Shi}, \bibinfo{author}{J.~Zhuang}, et~al.,
\newblock \bibinfo{title}{Imagetbad: A 3d computed tomography angiography image dataset for automatic segmentation of type-b aortic dissection},
\newblock \bibinfo{journal}{Frontiers in Physiology} \bibinfo{volume}{12} (\bibinfo{year}{2021}) \bibinfo{pages}{732711}.
\bibitem[{Wodzinski and Müller(2023)}]{r28}
\bibinfo{author}{M.~Wodzinski}, \bibinfo{author}{H.~Müller},
\newblock \bibinfo{title}{Automatic aorta segmentation with heavily augmented, high-resolution 3-d resunet: Contribution to the seg. a challenge},
\newblock \bibinfo{journal}{MICCAI Aorta Segm. Challenge}  (\bibinfo{year}{2023}) \bibinfo{pages}{42--54}.
\bibitem[{Xiong et~al.(2022)Xiong, Ding, Sun, Zhang, Guan, Zhang, Chen, Liu, Cheng, Zhao et~al.}]{r35}
\bibinfo{author}{X.~Xiong}, \bibinfo{author}{Y.~Ding}, \bibinfo{author}{C.~Sun}, \bibinfo{author}{Z.~Zhang}, \bibinfo{author}{X.~Guan}, \bibinfo{author}{T.~Zhang}, \bibinfo{author}{H.~Chen}, \bibinfo{author}{H.~Liu}, \bibinfo{author}{Z.~Cheng}, \bibinfo{author}{L.~Zhao}, et~al.,
\newblock \bibinfo{title}{A cascaded multi-task generative framework for detecting aortic dissection on 3-d non-contrast-enhanced computed tomography},
\newblock \bibinfo{journal}{IEEE Journal of Biomedical and Health Informatics} \bibinfo{volume}{26} (\bibinfo{year}{2022}) \bibinfo{pages}{5177--5188}.
\bibitem[{Sieren et~al.(2022)Sieren, Widmann, Weiss, Moltz, Link, Wegner, Stahlberg, Horn, Oecherting, Goltz et~al.}]{r36}
\bibinfo{author}{M.~M. Sieren}, \bibinfo{author}{C.~Widmann}, \bibinfo{author}{N.~Weiss}, \bibinfo{author}{J.~H. Moltz}, \bibinfo{author}{F.~Link}, \bibinfo{author}{F.~Wegner}, \bibinfo{author}{E.~Stahlberg}, \bibinfo{author}{M.~Horn}, \bibinfo{author}{T.~H. Oecherting}, \bibinfo{author}{J.~P. Goltz}, et~al.,
\newblock \bibinfo{title}{Automated segmentation and quantification of the healthy and diseased aorta in ct angiographies using a dedicated deep learning approach},
\newblock \bibinfo{journal}{European radiology} \bibinfo{volume}{32} (\bibinfo{year}{2022}) \bibinfo{pages}{690--701}.
\bibitem[{Li et~al.(2022)Li, Sun, Lam, Zhang, Sun, Peng, Xu, and Zhang}]{r37}
\bibinfo{author}{F.~Li}, \bibinfo{author}{L.~Sun}, \bibinfo{author}{K.-Y. Lam}, \bibinfo{author}{S.~Zhang}, \bibinfo{author}{Z.~Sun}, \bibinfo{author}{B.~Peng}, \bibinfo{author}{H.~Xu}, \bibinfo{author}{L.~Zhang},
\newblock \bibinfo{title}{Segmentation of human aorta using 3d nnu-net-oriented deep learning},
\newblock \bibinfo{journal}{Review of Scientific Instruments} \bibinfo{volume}{93} (\bibinfo{year}{2022}).
\bibitem[{Zhao and Feng(2021)}]{r38}
\bibinfo{author}{J.~Zhao}, \bibinfo{author}{Q.~Feng},
\newblock \bibinfo{title}{Automatic aortic dissection centerline extraction via morphology-guided crn tracker},
\newblock \bibinfo{journal}{IEEE Journal of Biomedical and Health Informatics} \bibinfo{volume}{25} (\bibinfo{year}{2021}) \bibinfo{pages}{3473--3485}.
\bibitem[{Feiger et~al.(2021)Feiger, Lorenzana-Saldivar, Cooke, Horstmeyer, Bishawi, Doberne, Hughes, Ranney, Voigt, and Randles}]{r39}
\bibinfo{author}{B.~Feiger}, \bibinfo{author}{E.~Lorenzana-Saldivar}, \bibinfo{author}{C.~Cooke}, \bibinfo{author}{R.~Horstmeyer}, \bibinfo{author}{M.~Bishawi}, \bibinfo{author}{J.~Doberne}, \bibinfo{author}{G.~C. Hughes}, \bibinfo{author}{D.~Ranney}, \bibinfo{author}{S.~Voigt}, \bibinfo{author}{A.~Randles},
\newblock \bibinfo{title}{Evaluation of u-net based architectures for automatic aortic dissection segmentation},
\newblock \bibinfo{journal}{ACM Transactions on Computing for Healthcare (HEALTH)} \bibinfo{volume}{3} (\bibinfo{year}{2021}) \bibinfo{pages}{1--16}.
\bibitem[{Wobben et~al.(2021)Wobben, Codari, Mistelbauer, Pepe, Higashigaito, Hahn, Mastrodicasa, Turner, Hinostroza, B{\"a}umler et~al.}]{r40}
\bibinfo{author}{L.~D. Wobben}, \bibinfo{author}{M.~Codari}, \bibinfo{author}{G.~Mistelbauer}, \bibinfo{author}{A.~Pepe}, \bibinfo{author}{K.~Higashigaito}, \bibinfo{author}{L.~D. Hahn}, \bibinfo{author}{D.~Mastrodicasa}, \bibinfo{author}{V.~L. Turner}, \bibinfo{author}{V.~Hinostroza}, \bibinfo{author}{K.~B{\"a}umler}, et~al.,
\newblock \bibinfo{title}{Deep learning-based 3d segmentation of true lumen, false lumen, and false lumen thrombosis in type-b aortic dissection},
\newblock in: \bibinfo{booktitle}{2021 43rd Annual International Conference of the IEEE Engineering in Medicine \& Biology Society (EMBC)}, \bibinfo{organization}{IEEE}, \bibinfo{year}{2021}, pp. \bibinfo{pages}{3912--3915}.
\bibitem[{Song et~al.(2022)Song, Chai, and Zhu}]{r41}
\bibinfo{author}{Z.~Song}, \bibinfo{author}{S.~Chai}, \bibinfo{author}{E.~Zhu},
\newblock \bibinfo{title}{Segmentation of aorta with aortic dissection based on centerline and boundary distance},
\newblock in: \bibinfo{booktitle}{2022 41st Chinese Control Conference (CCC)}, \bibinfo{organization}{IEEE}, \bibinfo{year}{2022}, pp. \bibinfo{pages}{7292--7297}.
\bibitem[{Hahn et~al.(2020)Hahn, Mistelbauer, Higashigaito, Koci, Willemink, Sailer, Fischbein, and Fleischmann}]{r42}
\bibinfo{author}{L.~D. Hahn}, \bibinfo{author}{G.~Mistelbauer}, \bibinfo{author}{K.~Higashigaito}, \bibinfo{author}{M.~Koci}, \bibinfo{author}{M.~J. Willemink}, \bibinfo{author}{A.~M. Sailer}, \bibinfo{author}{M.~Fischbein}, \bibinfo{author}{D.~Fleischmann},
\newblock \bibinfo{title}{Ct-based true-and false-lumen segmentation in type b aortic dissection using machine learning},
\newblock \bibinfo{journal}{Radiology: Cardiothoracic Imaging} \bibinfo{volume}{2} (\bibinfo{year}{2020}) \bibinfo{pages}{e190179}.
\bibitem[{Abdolmanafi et~al.(2023)Abdolmanafi, Forneris, Moore, and Di~Martino}]{r43}
\bibinfo{author}{A.~Abdolmanafi}, \bibinfo{author}{A.~Forneris}, \bibinfo{author}{R.~D. Moore}, \bibinfo{author}{E.~S. Di~Martino},
\newblock \bibinfo{title}{Deep-learning method for fully automatic segmentation of the abdominal aortic aneurysm from computed tomography imaging},
\newblock \bibinfo{journal}{Frontiers in Cardiovascular Medicine} \bibinfo{volume}{9} (\bibinfo{year}{2023}) \bibinfo{pages}{1040053}.
\bibitem[{Chen et~al.(2021)Chen, Zhang, Mei, Liao, Xu, Li, Xiao, Guo, Zhang, Yan et~al.}]{r44}
\bibinfo{author}{D.~Chen}, \bibinfo{author}{X.~Zhang}, \bibinfo{author}{Y.~Mei}, \bibinfo{author}{F.~Liao}, \bibinfo{author}{H.~Xu}, \bibinfo{author}{Z.~Li}, \bibinfo{author}{Q.~Xiao}, \bibinfo{author}{W.~Guo}, \bibinfo{author}{H.~Zhang}, \bibinfo{author}{T.~Yan}, et~al.,
\newblock \bibinfo{title}{Multi-stage learning for segmentation of aortic dissections using a prior aortic anatomy simplification},
\newblock \bibinfo{journal}{Medical image analysis} \bibinfo{volume}{69} (\bibinfo{year}{2021}) \bibinfo{pages}{101931}.
\bibitem[{Lyu et~al.(2021)Lyu, Yang, Zhao, Shu, Luo, Chen, Xiong, Yang, Li, Coatrieux et~al.}]{r45}
\bibinfo{author}{T.~Lyu}, \bibinfo{author}{G.~Yang}, \bibinfo{author}{X.~Zhao}, \bibinfo{author}{H.~Shu}, \bibinfo{author}{L.~Luo}, \bibinfo{author}{D.~Chen}, \bibinfo{author}{J.~Xiong}, \bibinfo{author}{J.~Yang}, \bibinfo{author}{S.~Li}, \bibinfo{author}{J.-L. Coatrieux}, et~al.,
\newblock \bibinfo{title}{Dissected aorta segmentation using convolutional neural networks},
\newblock \bibinfo{journal}{Computer methods and programs in biomedicine} \bibinfo{volume}{211} (\bibinfo{year}{2021}) \bibinfo{pages}{106417}.
\bibitem[{Zhao et~al.(2022)Zhao, Zhao, Pang, and Feng}]{r46}
\bibinfo{author}{J.~Zhao}, \bibinfo{author}{J.~Zhao}, \bibinfo{author}{S.~Pang}, \bibinfo{author}{Q.~Feng},
\newblock \bibinfo{title}{Segmentation of the true lumen of aorta dissection via morphology-constrained stepwise deep mesh regression},
\newblock \bibinfo{journal}{IEEE Transactions on Medical Imaging} \bibinfo{volume}{41} (\bibinfo{year}{2022}) \bibinfo{pages}{1826--1836}.
\bibitem[{Deng et~al.(2018)Deng, Zheng, Xu, Xi, Li, and Yin}]{r47}
\bibinfo{author}{X.~Deng}, \bibinfo{author}{Y.~Zheng}, \bibinfo{author}{Y.~Xu}, \bibinfo{author}{X.~Xi}, \bibinfo{author}{N.~Li}, \bibinfo{author}{Y.~Yin},
\newblock \bibinfo{title}{Graph cut based automatic aorta segmentation with an adaptive smoothness constraint in 3d abdominal ct images},
\newblock \bibinfo{journal}{Neurocomputing} \bibinfo{volume}{310} (\bibinfo{year}{2018}) \bibinfo{pages}{46--58}.
\bibitem[{Yu et~al.(2020)Yu, Gao, Wei, Liao, Xiao, Zhang, Yin, and Lu}]{r48}
\bibinfo{author}{Y.~Yu}, \bibinfo{author}{Y.~Gao}, \bibinfo{author}{J.~Wei}, \bibinfo{author}{F.~Liao}, \bibinfo{author}{Q.~Xiao}, \bibinfo{author}{J.~Zhang}, \bibinfo{author}{W.~Yin}, \bibinfo{author}{B.~Lu},
\newblock \bibinfo{title}{A three-dimensional deep convolutional neural network for automatic segmentation and diameter measurement of type b aortic dissection},
\newblock \bibinfo{journal}{Korean journal of radiology} \bibinfo{volume}{22} (\bibinfo{year}{2020}) \bibinfo{pages}{168}.
\bibitem[{Cheng et~al.(2020)Cheng, Tian, Yu, Ma, and Xing}]{r49}
\bibinfo{author}{J.~Cheng}, \bibinfo{author}{S.~Tian}, \bibinfo{author}{L.~Yu}, \bibinfo{author}{X.~Ma}, \bibinfo{author}{Y.~Xing},
\newblock \bibinfo{title}{A deep learning algorithm using contrast-enhanced computed tomography (ct) images for segmentation and rapid automatic detection of aortic dissection},
\newblock \bibinfo{journal}{Biomedical Signal Processing and Control} \bibinfo{volume}{62} (\bibinfo{year}{2020}) \bibinfo{pages}{102145}.
\bibitem[{Cao et~al.(2019)Cao, Shi, Ge, Xing, Zuo, Jia, Liu, He, Wang, Luan et~al.}]{r50}
\bibinfo{author}{L.~Cao}, \bibinfo{author}{R.~Shi}, \bibinfo{author}{Y.~Ge}, \bibinfo{author}{L.~Xing}, \bibinfo{author}{P.~Zuo}, \bibinfo{author}{Y.~Jia}, \bibinfo{author}{J.~Liu}, \bibinfo{author}{Y.~He}, \bibinfo{author}{X.~Wang}, \bibinfo{author}{S.~Luan}, et~al.,
\newblock \bibinfo{title}{Fully automatic segmentation of type b aortic dissection from cta images enabled by deep learning},
\newblock \bibinfo{journal}{European journal of radiology} \bibinfo{volume}{121} (\bibinfo{year}{2019}) \bibinfo{pages}{108713}.
\bibitem[{Ronneberger et~al.(2015)Ronneberger, Fischer, and Brox}]{r27}
\bibinfo{author}{O.~Ronneberger}, \bibinfo{author}{P.~Fischer}, \bibinfo{author}{T.~Brox},
\newblock \bibinfo{title}{U-net: Convolutional networks for biomedical image segmentation},
\newblock in: \bibinfo{booktitle}{Medical image computing and computer-assisted intervention--MICCAI 2015: 18th international conference, Munich, Germany}, \bibinfo{year}{2015}, pp. \bibinfo{pages}{234--241}.
\bibitem[{Chen et~al.(2021)Chen, Lu, Yu, Luo, Adeli, Wang, Lu, Yuille, and Zhou}]{r51}
\bibinfo{author}{J.~Chen}, \bibinfo{author}{Y.~Lu}, \bibinfo{author}{Q.~Yu}, \bibinfo{author}{X.~Luo}, \bibinfo{author}{E.~Adeli}, \bibinfo{author}{Y.~Wang}, \bibinfo{author}{L.~Lu}, \bibinfo{author}{A.~L. Yuille}, \bibinfo{author}{Y.~Zhou},
\newblock \bibinfo{title}{Transunet: Transformers make strong encoders for medical image segmentation},
\newblock \bibinfo{journal}{arXiv preprint arXiv:2102.04306}  (\bibinfo{year}{2021}).
\bibitem[{Zhao et~al.(2017)Zhao, Shi, Qi, Wang, and Jia}]{r52}
\bibinfo{author}{H.~Zhao}, \bibinfo{author}{J.~Shi}, \bibinfo{author}{X.~Qi}, \bibinfo{author}{X.~Wang}, \bibinfo{author}{J.~Jia},
\newblock \bibinfo{title}{Pyramid scene parsing network},
\newblock in: \bibinfo{booktitle}{Proceedings of the IEEE conference on computer vision and pattern recognition}, \bibinfo{year}{2017}, pp. \bibinfo{pages}{2881--2890}.
\bibitem[{Badrinarayanan et~al.(2017)Badrinarayanan, Kendall, and Cipolla}]{r53}
\bibinfo{author}{V.~Badrinarayanan}, \bibinfo{author}{A.~Kendall}, \bibinfo{author}{R.~Cipolla},
\newblock \bibinfo{title}{Segnet: A deep convolutional encoder-decoder architecture for image segmentation},
\newblock \bibinfo{journal}{IEEE transactions on pattern analysis and machine intelligence} \bibinfo{volume}{39} (\bibinfo{year}{2017}) \bibinfo{pages}{2481--2495}.
\bibitem[{Dosovitskiy et~al.(2020)Dosovitskiy, Beyer, Kolesnikov, Weissenborn, Zhai, Unterthiner, Dehghani, Minderer, Heigold, Gelly et~al.}]{r54}
\bibinfo{author}{A.~Dosovitskiy}, \bibinfo{author}{L.~Beyer}, \bibinfo{author}{A.~Kolesnikov}, \bibinfo{author}{D.~Weissenborn}, \bibinfo{author}{X.~Zhai}, \bibinfo{author}{T.~Unterthiner}, \bibinfo{author}{M.~Dehghani}, \bibinfo{author}{M.~Minderer}, \bibinfo{author}{G.~Heigold}, \bibinfo{author}{S.~Gelly}, et~al.,
\newblock \bibinfo{title}{An image is worth 16x16 words: Transformers for image recognition at scale},
\newblock \bibinfo{journal}{arXiv preprint arXiv:2010.11929}  (\bibinfo{year}{2020}).
\bibitem[{Liu et~al.(2021)Liu, Lin, Cao, Hu, Wei, Zhang, Lin, and Guo}]{r55}
\bibinfo{author}{Z.~Liu}, \bibinfo{author}{Y.~Lin}, \bibinfo{author}{Y.~Cao}, \bibinfo{author}{H.~Hu}, \bibinfo{author}{Y.~Wei}, \bibinfo{author}{Z.~Zhang}, \bibinfo{author}{S.~Lin}, \bibinfo{author}{B.~Guo},
\newblock \bibinfo{title}{Swin transformer: Hierarchical vision transformer using shifted windows},
\newblock in: \bibinfo{booktitle}{Proceedings of the IEEE/CVF international conference on computer vision}, \bibinfo{year}{2021}, pp. \bibinfo{pages}{10012--10022}.
\bibitem[{Xie et~al.(2021)Xie, Wang, Yu, Anandkumar, Alvarez, and Luo}]{r56}
\bibinfo{author}{E.~Xie}, \bibinfo{author}{W.~Wang}, \bibinfo{author}{Z.~Yu}, \bibinfo{author}{A.~Anandkumar}, \bibinfo{author}{J.~M. Alvarez}, \bibinfo{author}{P.~Luo},
\newblock \bibinfo{title}{Segformer: Simple and efficient design for semantic segmentation with transformers},
\newblock \bibinfo{journal}{Advances in neural information processing systems} \bibinfo{volume}{34} (\bibinfo{year}{2021}) \bibinfo{pages}{12077--12090}.

\end{thebibliography}

\end{document}